\documentclass[apjfonts,appendixfloats]{emulateapj}
\usepackage{appendix,natbib}
\usepackage{amsmath}
\usepackage{courier}
\usepackage{booktabs}
\usepackage{upgreek}
\usepackage[pdfpagemode=UseNone,pdfstartview=FitH,colorlinks=true,citecolor=blue,linkcolor=blue,urlcolor=blue]{hyperref}
\usepackage[all]{hypcap}

\newcommand{\hbeta}{\ensuremath{\mathrm{H} \beta}}
\newcommand{\kmps}{\ensuremath{\mathrm{km~s}^{-1}}}
\def\msun{{\rm\,M_\odot}}

\begin{document}

\defcitealias{rz14}{RZ14}
\defcitealias{rv13}{RV13}
\defcitealias{s16}{S16}

\title{KCWI observations of the extended nebulae in Mrk 273}

\author{\sc Gene C. K. Leung\altaffilmark{1,2}, 
Alison L. Coil\altaffilmark{1}, 
David S. N. Rupke\altaffilmark{3},
Serena Perrotta\altaffilmark{1}
}

\altaffiltext{1}{Center for Astrophysics and Space Sciences, University of California, San Diego, La Jolla, CA 92093, USA}
\altaffiltext{2}{Department of Astronomy, The University of Texas at Austin, Austin, TX 78712, USA}
\altaffiltext{3}{Department of Physics, Rhodes College, Memphis, TN 38112, USA}

\begin{abstract}

Ultraluminous infrared galaxies (ULIRGs) represent a critical stage in the merger-driven evolution of galaxies when AGN activity is common and AGN feedback is expected.
We present high sensitivity and large field of view intergral field spectroscopy of the ULIRG Mrk 273 using new data from the Keck Cosmic Web Imager (KWCI).
The KCWI data captures the complex nuclear region and the two extended nebulae in the northeast (NE) and southwest (SW) to $\sim 20$ kpc scales.
Kinematics in the nuclear region show a fast, extended, bipolar outflow in the direction of the previously reported nuclear superbubbles spanning $\sim 5$ kpc, two to three times greater than the previously reported size.
The larger scale extended nebulae on $\sim 20$ kpc show fairly uniform kinematics with FWHM $\sim 300 ~\kmps$ in the SW nebula and FWHM $\sim 120 ~\kmps$ in the NE nebula.
We detect for the first time high ionization [NeV]3426, [OIII]4363 and HeII4684 emission lines in the extended NE nebula.
Emission line ratios in the nuclear region correlate with the kinematic structures, with the bipolar outflow and the less collimated ``outflow regions'' showing distinct line ratio trends.
Line ratio diagnostics of high ionization emission lines reveal non-trivial contribution from shock plus precursor ionization in the NE nebula and the nuclear region, mixed with AGN photoionization.
These data are highly constraining for models of cool ionized gas existing ~20 kpc from a galactic nucleus.
\end{abstract}

\maketitle

\section{Introduction}

A critical unresolved problem in galaxy formation is the co-evolution of supermassive black holes (SMBHs) and their host galaxies.
Theoretical models of galaxy formation invoke feedback from active galactic nuclei (AGNs), by means of launching fast gas outflows, to 
quench star formation in the host galaxy and produce various observed galaxy properties \citep[e.g.][]{dm05, cro06, hop06a, kav17, phi18}.
In many evolutionary scenarios of galaxies and SMBHs, AGNs represent a key stage at the late stage of a galaxy merger in an evolutionary sequence \citep[e.g][]{san88,spr05,hop06a, sij07}.
In such scenarios, a major merger event triggers a rapid episode of gas inflow into the central region of the galaxy, fueling a burst of star formation activity and a dust-obscured AGN.
The AGN then drives outflows that couple with the surrounding interstellar medium and eventually removes the obscuring gas and dust in a ``blowout'', exposing the nucleus resulting in a luminous quasar.
These outflows are predicted to be the most powerful in the final stages of the merger, and are potentially sufficient to quench star formation in the environment \citep[e.g][]{deb12}.

Ultra-luminous infrared galaxies (ULIRGs) are characterized by their extreme far-infrared luminosities exceeding $10^{12} L_\odot$, which is believed to originate from dust heated by a starburst and/or AGN in the center of the galaxy \citep[e.g][]{san88,vei95,vei02}.
Moreover, morphological features consistent with major merging events, such as tidal tails or double nuclei, are almost ubiquitous among ULIRGs \citep[e.g][]{kim02,vei02}.
According to the merger-driven evolutionary scenarios, ULIRGs represent a population of galaxies at a critical evolutionary stage when the ``blowout'' is in action, and AGN-driven outflows are likely to be powerful. 
As such, ULIRGs provide an ideal laboratory to observationally test this evolutionary picture, and, in particular, to understand the role of AGN-driven outflows in the evolution of galaxies.

AGN-driven outflows are known to be prevalent in ULIRGs in the ionized (e.g. \citealt{har12}; \citealt{rz13}; \citealt{rv13}, hereafter \citetalias{rv13}, \citealt{arr14}), molecular \citep[e.g][]{cic12,rv13b,spo13,vei13,cic14,gon17,her20} and neutral (\citealt{rv11}, \citetalias{rv13}) phases.
The physical extent of the outflow is key to quantifying the impact of AGN-driven outflows in ULIRGs, as outflows extending to scales comparable to the size of the galaxy can have substantial impact on the star formation of the host galaxy, and the physical extent is required to determine the mass and energy outflow rate.
The physical extent of outflows remains uncertain despite their prevalence.
For ionized outflows, integral field unit (IFU) observations in the local Universe have revealed kiloparsec-scale ionized outflows in ULIRGs extending $\sim 2-12$ kpc in radius (e.g. \citealt{rv11}; \citetalias{rv13}; \citealt{rup17}), while at high redshift ($z\sim 2$), galaxy-wide outflows in ULIRGs have been observed to reach $\sim 4-14$ kpc \citep[e.g][]{ale10,har12}.
In contrast, recent long-slit and/or imaging studies of local ULIRGs have revealed more compact ionized outflows with sizes $\lesssim 1.5$ kpc \citep[e.g][]{ros18,tad18,spe18,tad19}.

Long-slit spectroscopy can potentially underestimate the size of the outflow, as spatial information is only available along the slit while collimated outflows \citepalias[e.g.][]{rv13} can be misaligned with the slit orientation.
IFU data can reveal the two-dimensional extent, however a critical uncertainty in IFU measurements of outflow extents at low redshift is often the limited field of view (FOV) of the instrument, as outflows often extend across much of the FOV \citepalias[e.g.][]{rv13}.
For example, the two-slit mode of the Gemini Multi-Object Spectrograph \citep[GMOS;][]{all02,hoo04} has a FOV of $5''\times 7''$, limiting the measurement of the projected galactocentric outflow extent to a radius of $\sim 2.5-3.5$ kpc at $z=0.05$.
Therefore, long-slit or FOV-limited IFU measurements of outflow extents in local ULIRGs are in principle lower limits and may underestimate the true extent.

The commissioning of the Keck Cosmic Web Imager \citep[KCWI;][]{mor18} on the Keck II Telescope presents new opportunities to advance the understanding of AGN-driven outflows in galaxy evolution with its large FOV and high throughput. 
KCWI is a general purpose optical IFU optimized for observations of faint, diffuse objects.
In the medium slicer configuration, KCWI has a FOV of $16'' \times 20''$, capable of mapping outflowing gas extending to a galactocentric radius of $8-10$ kpc at $z=0.05$, and covering an area over 9 times larger GMOS in one pointing.
KCWI also provides a high throughput, 3 to 20 times higher than GMOS, in the wavelength range spanning  $\sim 4000-5500~\AA$, which covers useful emission lines such as [OIII]5007, \hbeta, HeII4868 and [OIII]4363 at $z <0.1$.
Combined with the large field of view, KCWI is ideal for studying diffuse extended emission from AGN-driven outflows.
In this paper, we present a pilot study of a program to probe the true extent of AGN-driven outflows in low redshift galaxies with KCWI.

Mrk 273, one of the closest ULIRGs at $z=0.0373$, represents a unique target for this study.
It has a total star formation rate of $139~\msun ~\mathrm{yr~^{-1}}$ \citep{cic14}. 
It exhibits prominent tidal features which indicate it is a late-stage merger \citep[e.g][]{kim02}.
It has a complex nuclear structure hosting at least two to possibly three AGNs (see \citealt{rz14}, hereafter \citetalias{rz14}; \citealt{liu19} for recent discussions).
The two AGNs detected in X-ray are heavily obscured, with a combined bolometric luminosity of $\sim 5 \times 10^{44} ~\mathrm{erg~s^{-1}}$ \citep{iwa11,iwa18}.
Mrk 273 is known to host multi-phase high velocity AGN-driven outflows of various scales with a complex kinematic structure in the central $\sim 5$ kpc region.
On the sub-kiloparsec scale, compact molecular outflows have been observed with velocities of $>400~\kmps$ travelling towards the north up to $\sim 600$ pc from the nucleus \citep[e.g][]{u13,gon17,ala18}.
Long-slit spectroscopy and early IFU observations have revealed ionized outflows in the E-W direction about the nucleus with velocity widths (FWHM) $> 500~\kmps$ within a distance of $\sim 6$ kpc (\citealt{col99}; \citealt{rz13}; \citetalias{rz14}; \citealt{s16}, hereafter \citetalias{s16}).
IFU observation of the central $4.5 \times 6$ kpc shows a separate bipolar superbubble of fast outflowing ionized gas in the N-S direction extending 2 kpc on either side, reaching velocities of $\sim 1500~\kmps$ \citepalias{rv13}.

Apart from the fast ionized outflows in the central region, a striking feature of Mrk 273 are two extended nebulae of ionized gas stretching $>20$ kpc in the NE and SW directions, revealed by ground- and space-based imaging studies and long-slit spectroscopy \citepalias[\citealt{arm90};][]{rz14,s16}.
The nature of these extended nebulae is not completely clear. 
Given the presence of the multiple AGNs and fast central outflows in Mrk 273, it is of interest to determine the relation of the extended nebulae to the AGNs, such as whether they represent present or past episodes of outflows, or gas impacted by such outflows. 
Long-slit spectroscopy of selected apertures in the extended nebulae by \citet{xia99}, \citetalias{rz14} and \citetalias{s16} suggests relatively narrow kinematics, which can be associated with tidal debris, but the majority of the nebulae was not sampled.
The two extended nebulae are also detected in X-ray \citep{xia02,iwa11,liu19}.
\citet{liu19} argue that the high temperature and high $\alpha / \mathrm{Fe}$ ratio in the SW nebula suggests that it is a reservoir of gas accumulated from past episodes of outflows on a timescale of $\lesssim 0.1$ Gyr, where the turbulence in the gas has dissipated away.
IFU observations are required to robustly determine the nature of these extended nebulae, as they can map the full kinematic structure, and yield line ratio maps essential in understanding the physical conditions of the ionized gas.

In this paper, we analyze IFU observations of Mrk 273 using KCWI, with a focus on the nature of the extended nebulae.
In Section \ref{sec:data}, we describe the KCWI observations and the reduction and analysis of the data.
In Section \ref{sec:results}, we present our main results.
We discuss the implications of our results in Section \ref{sec:discussions} and summarize our findings in \ref{sec:conclusions}.
Throughout the paper we assume $H_0=67.4 ~\kmps ~\mathrm{Mpc}^{-1}$, $\Omega_\mathrm{m}=0.315$ and $\Omega_\mathrm{\Lambda} = 0.685$ \citep{pla20}.

\section{Observations and Data}\label{sec:data}

\subsection{KCWI observations}

\begin{figure}[!tp]\label{fig:fov}
	\centering
		\includegraphics[width=0.5\textwidth]{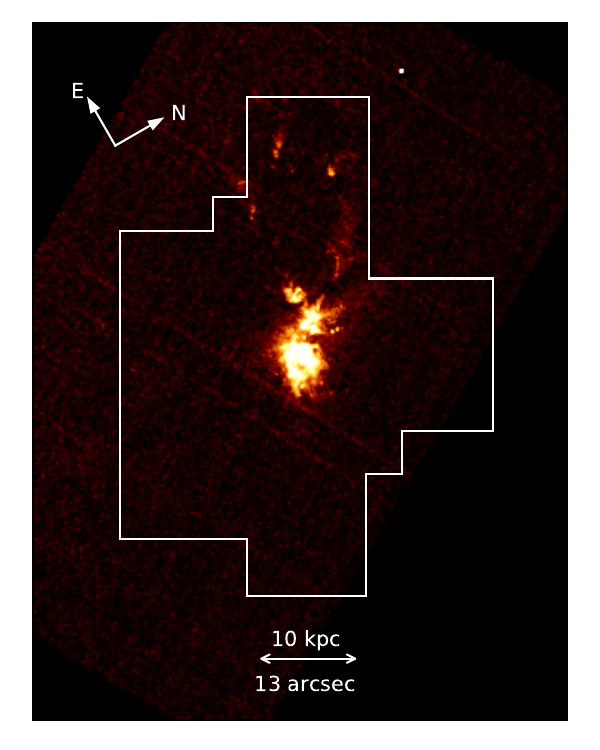}
		\caption{The outline of the footprint of our KCWI observation is shown in white, overlaid on the the HST/ACS [OIII] image in \citetalias{rz14}.}
\end{figure}

We observed Mrk 273 with KCWI on the Keck II telescope on June 14 and 15, 2018. 
We used the blue low-dispersion (BL) grating and medium slicer with KCWI, which provides a spectral resolution of $R = 1800$, a spaxel size of $0.29 \times 0.69 ''$, and a FOV of $20 '' \times 16 ''$ per pointing. 
We configured the articulating camera to a central wavelength of $4500 \AA$ and used a detector binning of $2 \times 2$, yielding wavelength coverage of $3435–5525 \AA$. 
The seeing was $0.8''$, corresponding to a projected distance of 0.62 kpc at the redshift of our target. 
Since the angular size of Mrk 273 exceeds the FOV of the IFU, we defined six pointing areas to cover the galaxy.
The footprint of our KCWI pointings are shown in Figure \ref{fig:fov}, overlaid on the HST/ACS [OIII] image in \citetalias{rz14}.
Each pointing area was observed for 30 to 60 minutes, consisting of individual exposures of 10 minutes for the nuclear region, and 15 minutes for the extended regions.
Individual exposures in each pointing area were dithered $0.35 ''$ along slices to subsample the output spaxels. 
For the nuclear region, we took a 1-minute exposure for each dither position to sample pixels saturated in the 10-minute exposure.

\subsection{Data reduction}

We reduced the data using the KCWI Data Extraction and Reduction Pipeline v1.1.0 and the IFSRED library \citep{rup14r}.
We first performed a  correction to saturated pixels in raw frames of the nuclear region using the routine IFSR\_KCWISATCOR.
This routine searches for saturated pixels in the 10-minute exposure of the nuclear region, and replaces them with pixels in its corresponding 1-minute exposure after scaling for the exposure time difference.
The corrected frames of the nuclear pointing region and raw frames of other pointing regions are then passed to the pipeline.
The default wavelength calibration produced large residuals of $\approx 1 \AA$, resulting from a mismatch with the pipeline thorium-argon (ThAr) atlas, so we selected DISPLAY mode 3 to manually accept and reject lines matched by the pipeline.
The residuals of the resulting wavelength solutions were reduced to $\approx 0.3 \AA$.
Sky subtraction was performed using the pipeline with manual selection of a sky mask region in each frame.
For pointings in the nuclear region, where the source occupies the entire FOV, we used a sky frame from a nearby pointing that was observed immediately before or after the nuclear pointing.
Following the full pipeline stages, we resampled the data onto $0.29'' \times 0.29''$ spaxel grids using the routine IFSR\_KCWIRESAMPLE, and generated a mosaic of the data using the routine IFSR\_MOSAIC.
The resulting stacked data cube has dimensions of $166 \times 215$ spaxels, covering $48.1'' \times 62.4''$ and corresponding to projected physical dimensions of $37.1~\mathrm{kpc} \times 48.1~\mathrm{kpc}$ at $z=0.0373$.

\subsection{Data Analysis}

\begin{figure}[!tp]\label{fig:vor}
	\centering
		\includegraphics[width=0.48\textwidth]{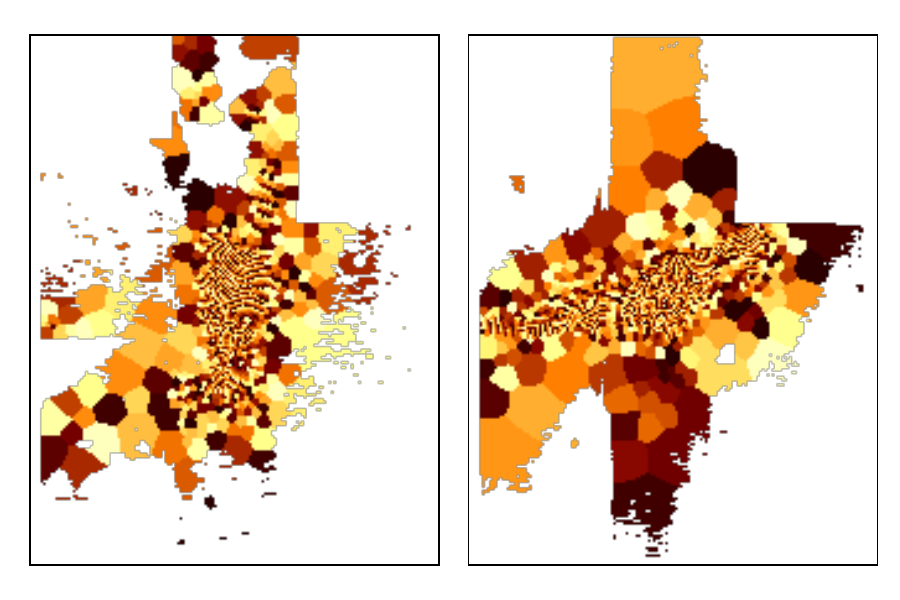}
		\caption{Voronoi binning schemes used for line (left) and continuum (right) emission in the KCWI data of Mrk 273.}
\end{figure}

\subsubsection{Voronoi binning}\label{subsec:vor}

To enhance the  signal-to-noise (S/N) for spectroscopic analysis of low surface brightness diffuse emission, we constructed Voronoi bins \citep{cap03} of the data using the IDL routine VORONOI\_2D\_BINNING \footnote{\url{https://www-astro.physics.ox.ac.uk/~mxc/software/\#binning}}.
The Voronoi binning technique is an optimal method to bin two-dimensional data to achieve a minimum S/N in each bin while maximally preserving spatial resolution, and is commonly used in the analysis of IFU data \citep{cap09}.
An emission line flux and error map is required to construct the Voronoi bins. 
The two strongest emission line features in our data are the [OII]3726,9 doublet and [OIII]5007.
We created initial emission line maps of each line for the purpose of Voronoi binning by integrating over the emission line and subtracting nearby continuum at wavelengths on either side.
The relative emission line strengths of [OIII]5007 and [OII]3726,9 vary widely across different regions of the galaxy.
For example, the [OIII]5007 emission line is several times stronger than the [OII]3726,9 doublet in the northeast region of the galaxy, but is several times weaker in the southwest. 
To minimize binning bias towards one particular emission line, and thus a particular region of the galaxy, we created a combined line map of both emission lines by multiplying the [OII]3726,9 map by two and adding it to the [OIII]5007 map.
The factor of two was introduced to avoid bias towards [OIII]5007, which is, on average, approximately two times stronger than [OII]3726,9 across the galaxy. 
We used the combined line flux and error maps to construct Voronoi bins with a target S/N of 25 and a threshold S/N of 1.
The 12184 spaxels above the S/N threshold are grouped into 1588 Voronoi bins, with 1026 unbinned spaxels. 
The resulting binning scheme is shown in the left panel of Figure \ref{fig:vor}.

\begin{figure*}[!htbp]\label{fig:image}
	\centering
		\includegraphics[width=\textwidth]{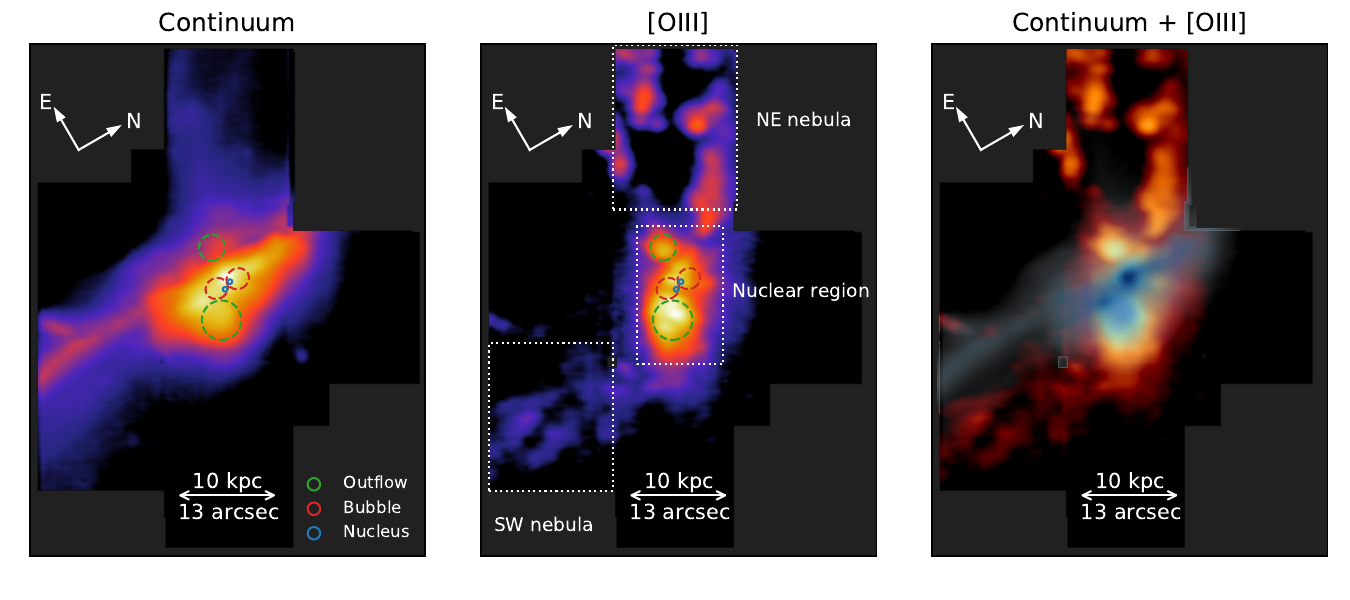}
		\caption{{\em Left}: Continuum image of Mrk 273 showing stellar emission. A tidal tail extends to the south of the galaxy. {\em Middle}: Continuum-subtracted [OIII]4959+5007 image showing emission of ionized gas. Two extended nebulae are observed in the NE and SW directions. The NE nebula contains several bright regions around its edge, and is hollow at the center. {\em Right}: Composite image of [OIII]4959+5007 in red and the continuum in blue. In the left and middle panels, the green-dashed circles denote the locations of the ``outflow regions'' reported in \citetalias{rz14}, the red-dashed circles denote location of the bipolar superbubble reported in \citetalias{rv13}, and the blue circles denote the two AGNs with X-ray detections. In the middle panel, the white-dashed rectangles indicate the regions used in our line ratio analysis for the extended nebulae and the nuclear region in Section \ref{subsec:ion}.}
\end{figure*}

This Voronoi-binned data cube constructed from bright emission lines traces the distribution of ionized gas in the galaxy, but it does not necessarily trace the distribution of the stellar continuum.
For example, Voronoi bins in regions with strong line emission but weak continuum emission will have a high S/N in emission lines but low S/N in the continuum. 
While this is likely sufficient for obtaining an accurate continuum flux level near emission lines to measure emission line fluxes and kinematics, other continuum parameters such as stellar kinematics may potentially have suboptimal accuracy.
Therefore, for the purpose of measuring continuum kinematics only, we constructed a separate set of Voronoi bins using flux and error maps of the continuum integrated between 3600 \AA ~and 5000 \AA ~with emission lines masked, and a target integrated S/N of 100 and a threshold integrated S/N of 3.
A total of 14664 spaxels were above the S/N threshold, among which 1875 Voronoi bins were constructed, with 1204 unbinned spaxels.
The resulting binning scheme is shown in the right panel of Figure \ref{fig:vor}.

\subsubsection{Emission-line fitting}\label{subsec:emlfit}
We use the combined [OII] and [OIII] Voronoi-binned data cube for emission line fitting.
We modelled the spectrum in each Voronoi bin using the IFSFIT library \citep{rup14f} in IDL.
It incorporates PPXF \citep{cap12} to fit the stellar continuum, and MPFIT \citep{mar09} to fit a user-defined number of Gaussian profiles to the emission lines.
IFSFIT first masks emission line regions and fits the stellar continuum, then simultaneously fits the emission lines in the continuum-subtracted spectrum. 
The continuum spectrum is fitted with the \citet{gon05} high-resolution stellar population synthesis model assuming solar metallicity, and Legendre polynomials to account for residuals from imperfect calibration such as scattered-light or residual sky background.
A maximum of three Gaussian components are allowed in each spaxel for each emisison line, where the kinematics of each component is tied to that of [OIII]5007.
Model line profiles were convolved with the spectral resolution before fitting.
The [OII] doublet is unresolved in our observation, and the [OII]3729/3726 flux ratio is fixed to 1.2, corresponding to an electron density of 400 cm$^{-3}$ \citep{pra06}.
To determine the number of components needed in each spaxel, the spectrum is first fitted with three Gaussian components. 
A component is only kept if it reaches $3\sigma$ significance in at least one strong emission line; otherwise, the component is removed and the spectrum is re-fitted.
After fitting with the final number of components, emission lines with a significance of less than $3\sigma$ in the total flux are set to zero.
From this we created flux and kinematics maps for each emission line.

To measure stellar kinematics, we additionally performed a continuum-only fit to the continuum Voronoi-binned data cube.
Emission lines were masked and the continuum was fitted with the same SPS model and Legendre polynomials described above.

\section{Results}\label{sec:results}

\subsection{Morphology}

\begin{figure*}[!htbp]\label{fig:lines_image}
	\centering
		\includegraphics[width=\textwidth]{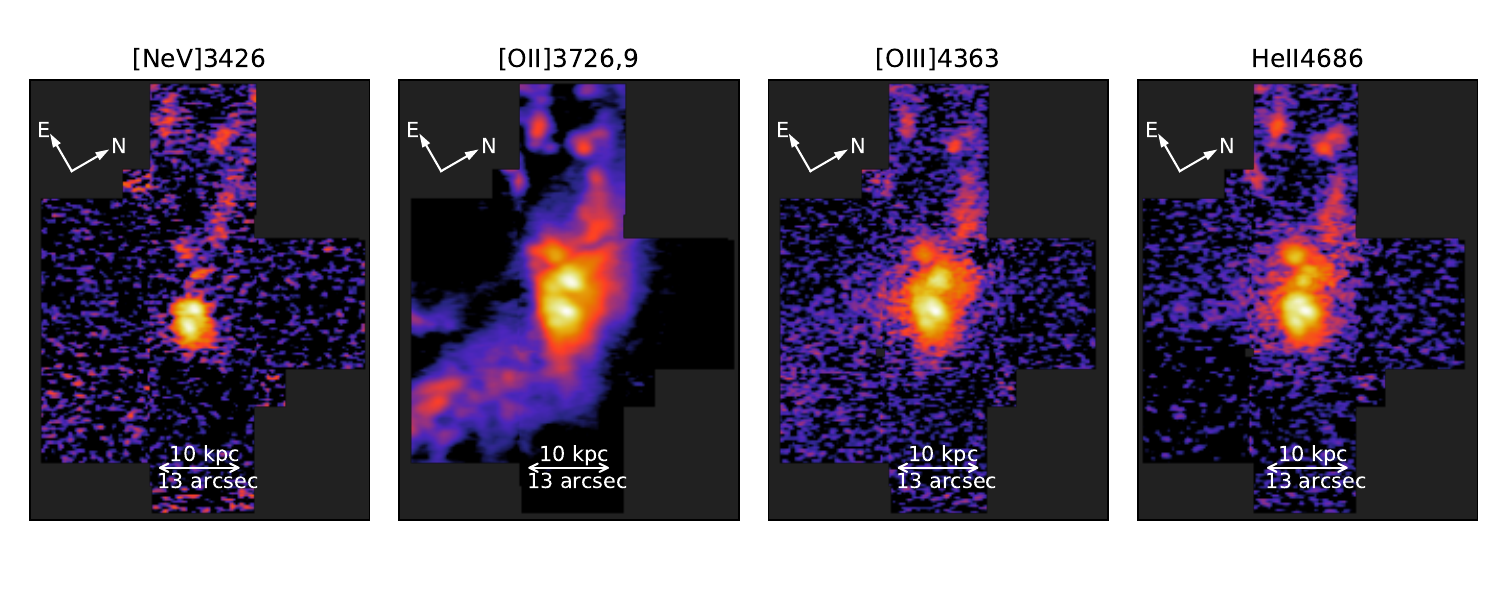}
		\caption{From left to right: Continuum-subtracted images of [NeV]3426, [OII]3726,9, [OIII]4363 and HeII4684. High ionization line emission in [NeV]3426, [OIII]4363 and HeII4684 is observed in not only the nuclear region, but also the NE nebula.}
\end{figure*}

KCWI has a remarkable capability to map low surface brightness extended emission.
The left panel of Figure \ref{fig:image} shows the mosaic image of the stellar continuum emission at $3600-5000\AA$ (wavelengths near emission lines have been  masked).
A prominent feature is the well-known tidal tail (e.g. \citealt{kim02}; \citetalias{rz14, s16}) extending to the south of the galaxy, which is common in galaxies in the late stage of a merger event\footnote{See Figure 5 of \citet{cox08} for the simulation of a merger between a galaxy pair with a stellar ratio of 10:1 at 3.6--4 Gyr with a strikingly similar morphology.}.
This image covers $\sim 12$ kpc of the tidal tail; the full extent is up to $\sim 36$ kpc \citep[e.g][]{kim02}.
We also observe  weak diffuse continuum emission in the northeast of the galaxy reported in \citetalias{rz14}. 

The middle panel of Figure \ref{fig:image} shows the continuum-subtracted [OIII]4959+5007 mosaic image.
The emission image was constructed by integrating $\pm 675$ \kmps around the doublet and subtracting nearby continuum on both sides.\footnote{For clarity, we note that the emission line maps presented in this Section are different from the initial line maps used to generate Voronoi bins.}
The brightest [OIII] emission is in the central 10 kpc of the galaxy, which we call the nuclear region in this paper.
In the Figure, we mark the locations of the two known AGNs, the nuclear superbubble in \citetalias{rv13} and the ``outflow regions'' in \citetalias{rz14}.
The nuclear superbubble extends in the N-S directions about the nuclei, with the northern half coinciding with a relatively bright region in [OIII], and the southern half  with a relatively faint region in [OIII].
This is consistent with the effect of extinction and the superbubble being a bipolar outflow, with blueshifted gas in the north and redshifted gas in the south (see Section \ref{sec:kin} for details).
The ``outflow regions'' defined by \citetalias{rz14} lie ENE and WSW from the nucleus, corresponding to two bright regions in [OIII].
We note that these outflows regions are constructed in \citetalias{rz14} to enclose regions of enhanced [OIII] emission immediately outside the nucleus, and their locations should not necessarily be taken as the direction of an outflow; we discuss this point further in Section \ref{sec:oflw}.

\begin{figure*}[!htbp]\label{fig:map_oiii}
	\centering
		\includegraphics[width=\textwidth]{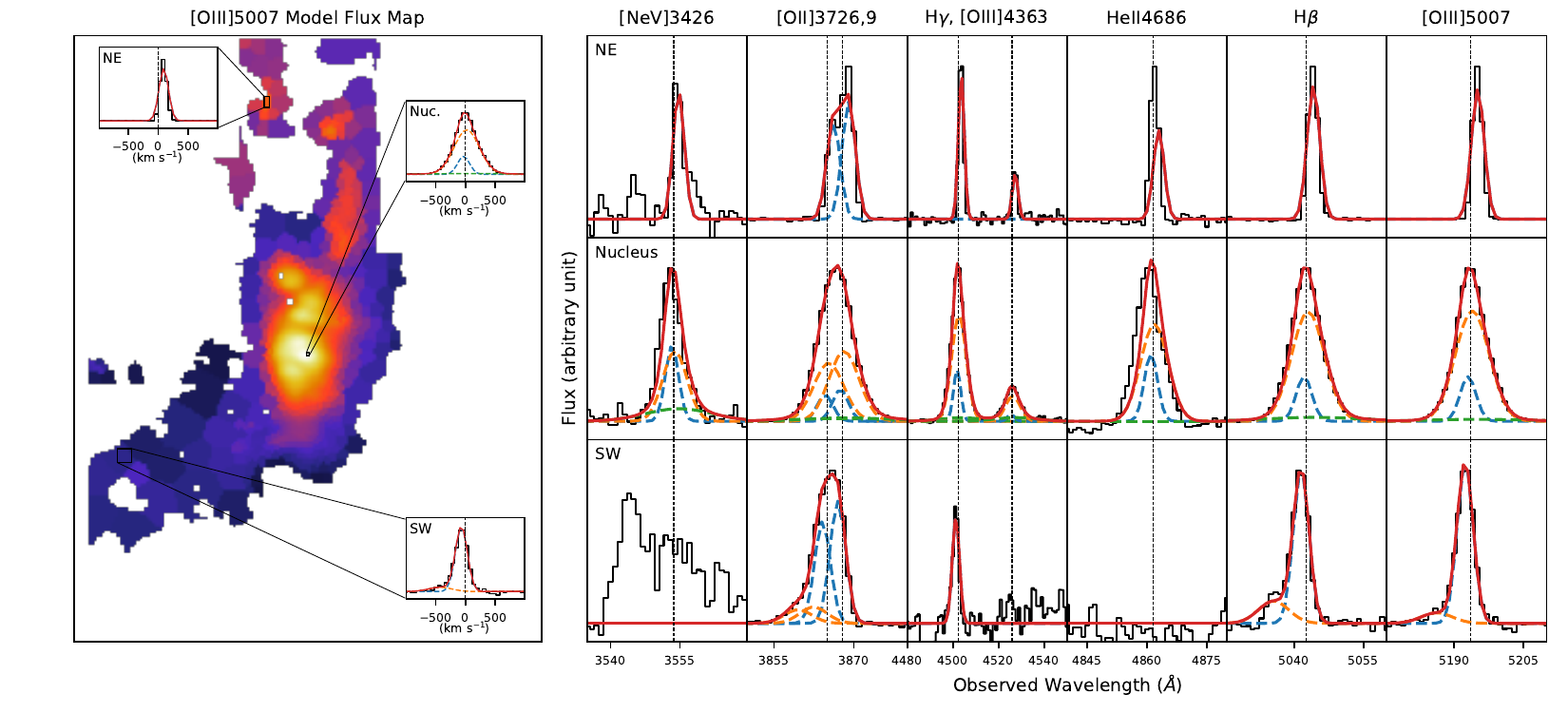}
		\caption{{\em Left}: Voronoi-binned model flux map of [OIII]5007. [OIII]5007 velocity profiles of representative Voronoi bins in the NE nebula, nuclear region and SW nebula are shown in the insets. {\em Right}: Spectra of each representative Voronoi bin from top to bottom. The emission lines are fitted with a maximum of three kinematic components. In the velocity profiles and spectra, the black line is the continuum-subtracted spectrum, the red line is the total emission line model, the blue, orange and green dashed lines are the emission line models for individual kinematic components. The nuclear bin has the broadest velocity profile, followed by the SW bin, while the NE bin has the narrowest profile. The SW exposures are affected by systematic features resulting in broad bumps in the spectra in the proximity of 3500 \AA . The [NeV]3426 emission line is therefore undetected in that region. See Section \ref{sec:eml} for details.}
\end{figure*}

We also detect in [OIII] the two previously-reported extended ~nebulae (\citealt{arm90}; \citetalias{rz14, s16}).
Extending from the ENE ``outflow region'', a bright nebula of ionized gas is seen spanning at least 20 kpc from the nucleus in the northeast of the galaxy.
This NE nebula contains several bright filamentary structures around its edge, and is hollow at the center, resembling an evacuated bubble or a loop.
While it is in the general direction of the diffuse continuum emission, the line and stellar emission regions in this direction do not completely overlap, particularly in the northern part of the nebula.
Another extended nebula is detected in [OIII] for the first time towards the southwest of the galaxy, west of the tidal tail.
This SW nebula extends $\sim 20$ kpc from the nucleus. 
It is less luminous in [OIII] compared with the NE nebula.
It approximately follows the direction of the tidal tail, but they are offset by $\sim 1-3$ kpc.
A composite continuum and [OIII] image is shown in the right panel of Figure \ref{fig:image} for comparison.

\subsection{Emission lines}\label{sec:eml}

\begin{figure*}[!htbp]\label{fig:map_lines}
	\centering
		\includegraphics[width=\textwidth]{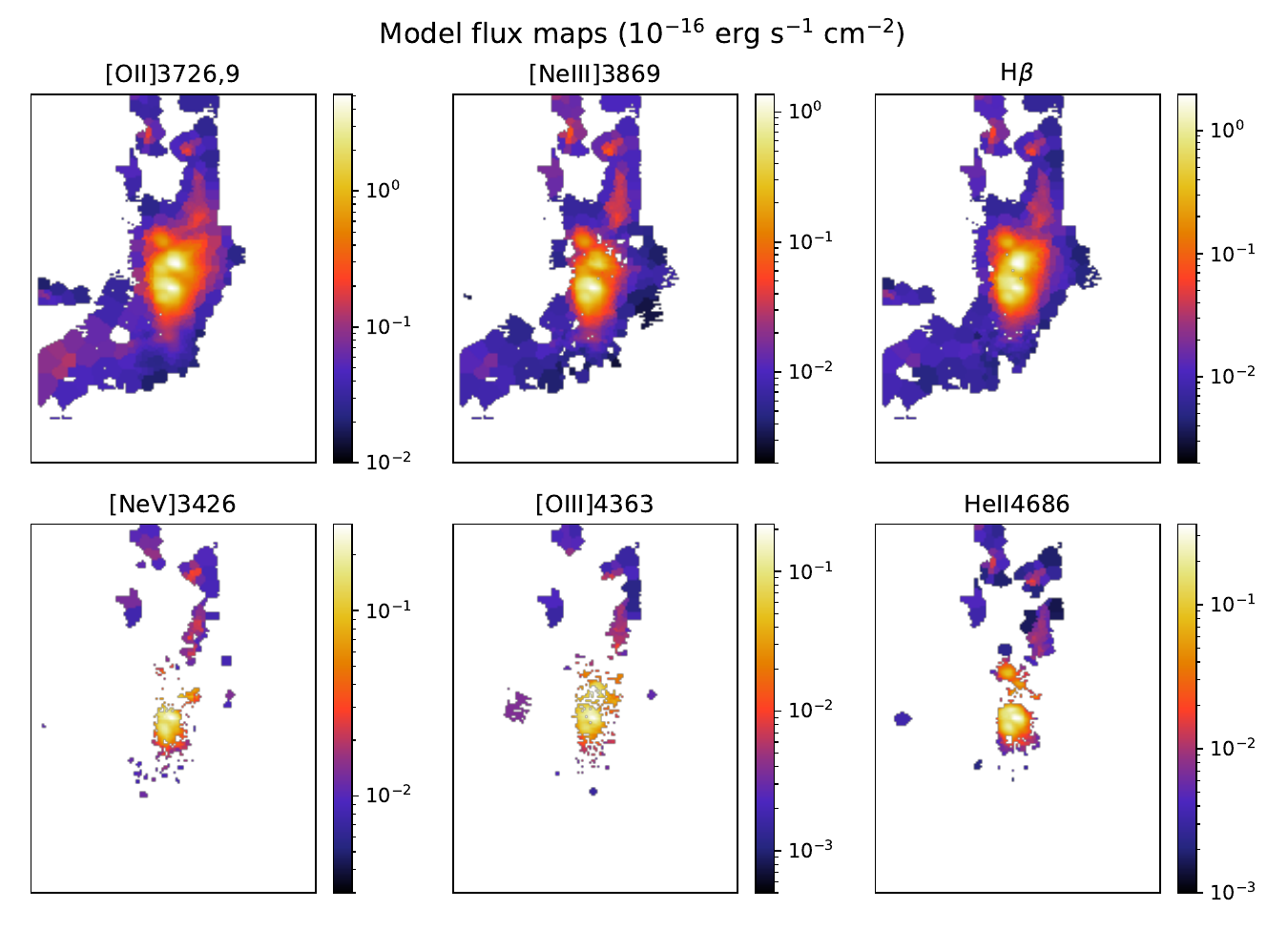}
		\caption{Voronoi-binned model flux map of [OII]3726,9, [NeIII]3869, \hbeta , [NeV]3426, [OIII]4363 and HeII4684. The color bars show the flux in units of $10^{-16} ~\mathrm{erg~s^{-1}~cm^{-2}}$. The high ionization [NeV]3426, [OIII]4363 and HeII4684 lines are detected in the NE nebula.}
\end{figure*}

Our wide-field IFU data allows us to map the flux and spatial distribution of numerous emission lines apart from the [OIII] doublet detected with KCWI.
A key result of this paper is the detection and spatial mapping of the faint, high-ionization emission lines [NeV]3426, [OIII]4363 and [HeII]4686, which are useful as shock and AGN diagnostics.
In addition, we map the emission from the low-ionization [OII]3726,9 doublet.
Figure \ref{fig:lines_image} shows the continuum-subtracted mosaic images of these emisison lines, integrating $\pm 675$ \kmps ~around the respective line centers at $z=0.0373$.
The high-ionization [NeV]3426, [OIII]4363 and HeII4686 lines show the strongest emission in the nuclear region.
Remarkably, these high-ionization lines are all observed to have extended emission in the NE nebula, in regions tracing the emission of the [OIII] doublet, but they are not detected in the SW nebula.
On the other hand, the morphology of the low-ionization [OII] doublet emission is similar to that of the [OIII] doublet, but it is enhanced in the SW nebula.

Quantitative measurements of emission line fluxes are obtained using the emission-line fitting procedures on the Voronoi-binned data cube described in Section \ref{subsec:emlfit}.
The left panel of Figure \ref{fig:map_oiii} shows the resulting Voronoi-binned model flux map of [OIII]5007, and the velocity profiles of three representative Voronoi bins from the nuclear region, NE nebula, and SW nebula.
The nuclear bin has the broadest velocity profile, followed by the SW bin, while the NE bin has the narrowest profile.
The detailed kinematic properties of the ionized gas are discussed below in Section \ref{sec:kin}.
In the right panel of Figure \ref{fig:map_oiii} we show the emission line spectra and best-fit multi-component models for the [NeV]3426, [OII]3726,9, H$\gamma$, [OIII]4363, HeII4686, \hbeta , and [OIII]5007 emission lines.
The [OII] doublet, Balmer lines, and [OIII]5007 emission lines are detected in all three spatial bins shown in the left panel, while the higher-ionization [NeV]3426, [OIII]4363 and HeII4686 are only detected in the nuclear and NE bins, and not in the SW bin.
The above properties are generally true for most Voronoi bins in each region.
Three kinematic components are typically required in the nuclear region, one to two in the SW nebula, and one in the NE nebula.

In the SW region, a series of broad bumps accompanied by large flux variances are observed in the spectra near the bluest end of the wavelength coverage of our observation, possibly due to imperfect calibration or differential observing conditions during the exposures in this pointing.
The [NeV]3426 emission line, which is within the wavelengths affected, is not detected in the SW bin.
Two other high-ionization lines, [OIII]4363 and HeII4686, have lower ionization potentials than [NeV]3426 and are thus potentially more prominent.
However, they are also undetected in the SW bin, 
even though they are not within the affected wavelengths. 
This forms a consistent picture that high-ionization lines are absent in the SW region.
Therefore, the non-detection of the [NeV]3426 emission line in the SW bin is likely due to an intrinsic absence of the line rather than the large flux variance observed in that region.

Figure \ref{fig:map_lines} presents a more complete spatial view of each emission line 
by showing the model flux maps of the [OII]3726,9 doublet, and the [NeIII]3869, \hbeta , [NeV]3426, [OIII]4363 and HeII4686 emission lines, using the emission line fits above. 
The [OII]3726,9 doublet, and the [NeIII]3869 and \hbeta ~emission lines are generally detected in the nuclear, NE and SW regions.
However, the high-ionization [NeV]3426, [OIII]4363 and HeII4686 emission lines are detected in only the nuclear and NE regions, and not in the SW region.

\begin{figure*}[!htbp]\label{fig:kin}
	\centering
		\includegraphics[width=0.7\textwidth]{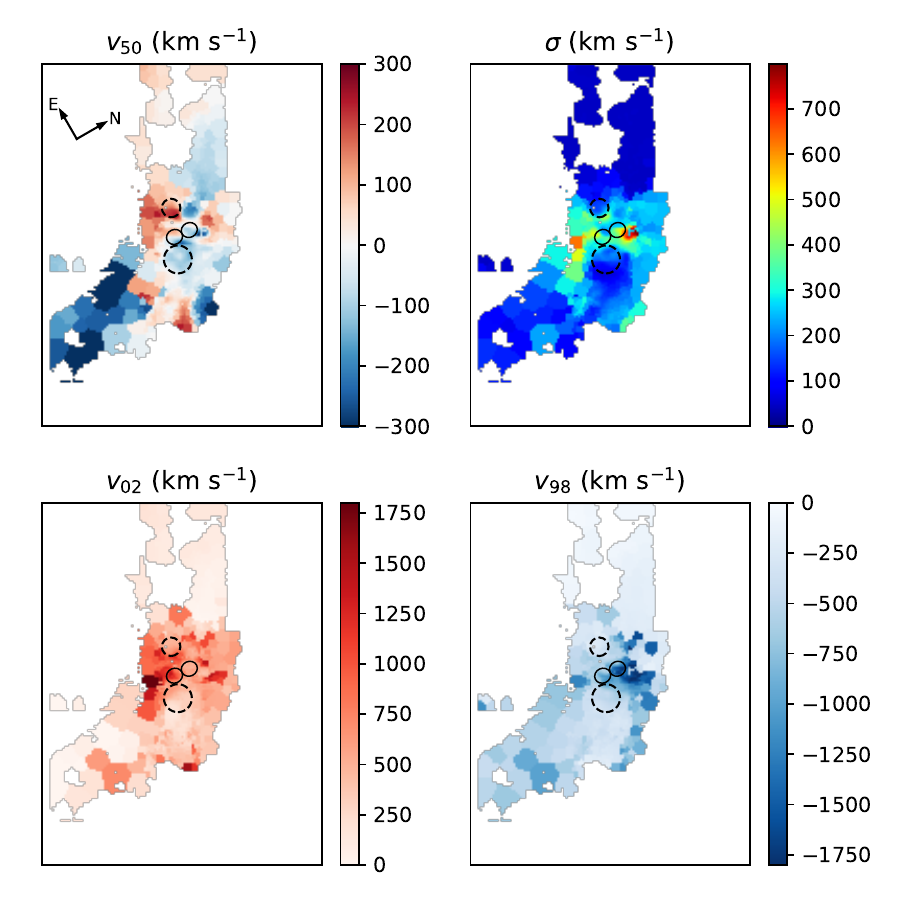}
		\caption{The central velocity ($v_{50}$), velocity dispersion ($\sigma$), maximum redshifted velocity ($v_{02}$) and maximum blueshifted velocity ($v_{98}$) of [OIII]5007. The color bars show the velocity in \kmps . The black dashed circles show the location of the ``outflow regions'', and the black solid circles show the location of the ``supperbubble''. Regions of high $\sigma > 250 ~\kmps$ are observed along the direction of the bipolar superbubbles, but extend to $\sim 5$ kpc, approximately two to three times the previously-reported size. Moderately broad emission with $\sigma \sim 150-250 ~\kmps$ is observed in the ``outflow regions'' extending $\sim 5$ kpc in the NE and SW directions. The larger scale extended nebulae have fairly uniform velocity dispersion, $\sigma \sim 100-150 ~\kmps$ in the SW nebula and $\sigma \sim 50 ~\kmps$ in the NE nebula.}
\end{figure*}

\subsection{Kinematics}\label{sec:kin}

With the integral field KCWI data, we are able to map out the kinematics of the ionized gas in the galaxy and identify distinct kinematic structures within the system.
An important result of this paper is the kinematics of the large scale ionized gas, i.e. the NE and SW nebulae.
The velocity percentiles presented below are with respect to the total line profile resulting from the sum of all significant components.
The upper left panel of Figure \ref{fig:kin} shows the central velocity $v_{50}$, which is the 50th percentile of the velocity profile, traced by [OIII]5007.
The systemic velocity is set to be the median of the central velocity of the stellar continuum.
The nuclear region has a complex kinematic structure, with a redshifted region towards the southeast and north of the nucleus, and a blueshifted region towards the west and northeast.
The large-scale SW nebula is generally blueshifted by $\sim 200-300$ \kmps.
The NE nebula has a mild redshift of $\sim 30-100$ \kmps ~in the eastern part of the nebula (seen as the left half in the figure), and is mildly blueshifted by $\sim 40-100$ \kmps ~in the western part (seen as the right half in the figure).

The upper right panel of Figure \ref{fig:kin} shows the velocity dispersion $\sigma$ of [OIII]5007 across the galaxy.
In the nuclear region, very broad emission with $\sigma > 250$ \kmps ~is observed along the direction of the bipolar superbubbles in \citetalias{rv13} (marked here as open black circles outlined with a solid line), but extends to $\sim 5$ kpc, approximately two to three times the previously-reported size of the superbubbles. 
The regions with the highest velocity dispersion ($\gtrsim 600 \kmps$) lie beyond the known extent of the superbubbles in the north and east directions. 
This suggests that the high-velocity outflowing gas in the superbubbles is substantially more extended that previously measured.
Moderately broad emission with $\sigma \sim 150-250$ \kmps ~is observed in the ``outflow regions'' defined in \citetalias{rz14}, which extend $\sim 5$ kpc in both directions perpendicular to the superbubbles (marked here with open circles with dashed outlines). 
The larger scale extended nebulae have fairly uniform velocity dispersion, with the SW nebula showing $\sigma \sim 100-150$ \kmps and the NE nebula showing a smaller dispersion of $\sigma \sim 50$ \kmps.

The lower left and lower right panels of Figure \ref{fig:kin} show the ``maximum'' redshifted and blueshifted velocities $v_{02}$ and $v_{98}$, which are the velocity that encompasses $2\%$ and $98\%$ of the cumulative velocity distribution, respectively.
The southern superbubble contains highly redshifted emission with $v_{02}$ between 800 and 1400 \kmps, while the northern superbubble contains highly blueshifted emission with $v_{98}$ bewteen $-1800$ and $-800$ \kmps , consistent with a bipolar superbubble of fast moving outflows.
Additionally, fast moving, redshifted gas with $v_{02} \gtrsim 900$ \kmps ~extends beyond the known superbubble towards a wedge-shaped region in the south up to $\sim 5$ kpc. 
Similarly, there is a wedge-shaped region of high-velocity blueshifted gas extending north up to $\sim 4$ kpc.
This, combined with the velocity dispersion map, suggests that fast moving gas from the superbubbles extends two to three times beyond the previously-known extent. On larger scales, the NE nebula has ``maximum'' velocities up to $-200$ and 200 \kmps , while the SW nebula generally has velocities between $-50$ and $500$ \kmps.

\begin{figure*}[!htbp]\label{fig:cont}
	\centering
		\includegraphics[width=\textwidth]{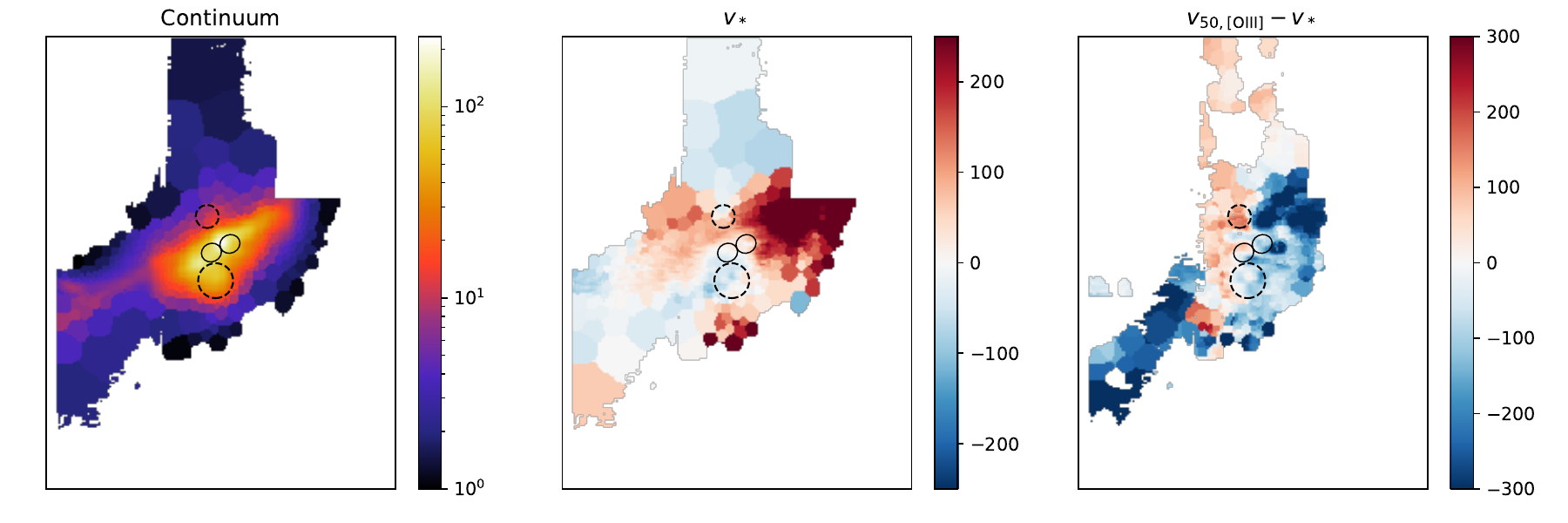}
		\caption{{\em Left}: Voronoi-binned model flux map for the contiuum. The color bar shows flux in units of $10^{-16} ~\mathrm{erg~s^{-1}~cm^{-2}}$. {\em Middle}: Stellar velocity map. {\em Right}: Map of [OIII]5007 emission central velocity relative to stellar velocity. The color bars of the middle and right panels show the velocity in \kmps . The black dashed circles show the location of the ``outflow regions'', and the black solid circles show the location of the supperbubbles. }
\end{figure*}

The kinematics of the stellar emission is measured by fitting the continuum Voronoi-binned data cube described in Section \ref{subsec:vor}.
The left panel of Figure \ref{fig:cont} shows the resulting model flux map of the continuum, while the middle panel shows the central velocity of the stellar continuum $v_*$.
On large scales, apart from the tidal tail,  weak continuum in the NE is also detected.
The stellar continuum near the nucleus is redshifted in the north-south direction, and blueshifted in the east-west direction.
In addition, the weak continuum in the NE is generally blueshifted by $\sim 100$ \kmps.
To quantify how much the kinematics of the ionized gas decouples from the continuum, 
we calculate the central velocity of [OIII]5007 relative to that of the stellar continuum by subtracting spaxel-by-spaxel the $v_*$ map of the continuum from the $v_{50}$ map of [OIII]5007.
We show the resulting map of $v_{50} - v_*$ in the the right panel of Figure \ref{fig:cont}.
The ionized gas north of the nucleus is generally blueshifted relative to the stellar continuum, and generally redshifted in the south, despite the complex kinematic structure in [OIII]5007 alone.
In addition, the NE nebula is generally redshifted from the stellar continuum by $\sim 50-100$ \kmps , instead of displaying the two distinct redshifted and blueshifted halves as seen in [OIII]5007 alone.

\begin{figure}[!tbp]\label{fig:ebv}
	\centering
		\includegraphics[width=0.4\textwidth]{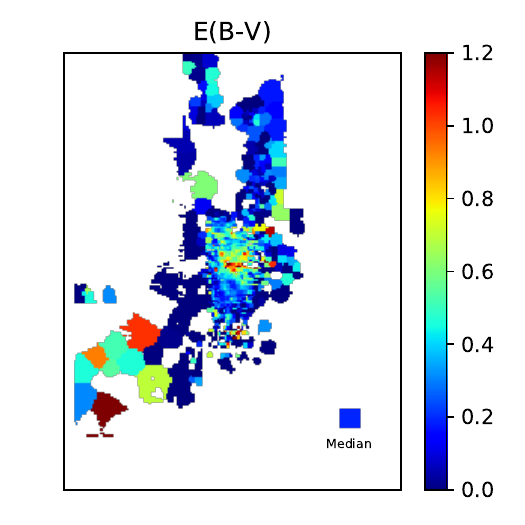}
		\caption{Extinction map computed from the Balmer decrement of the \hbeta ~and $H\gamma$ emission lines. The square shows the median extinction value of 0.19. Relatively strong reddening of $E(B-V) \gtrsim 0.5$ is observed in the the SW nebula as well as the inner nuclear region, while the NE nebula has weaker reddening of $\sim 0.2$.}
\end{figure}

\begin{figure*}[!htbp]\label{fig:lr}
	\centering
		\includegraphics[width=\textwidth]{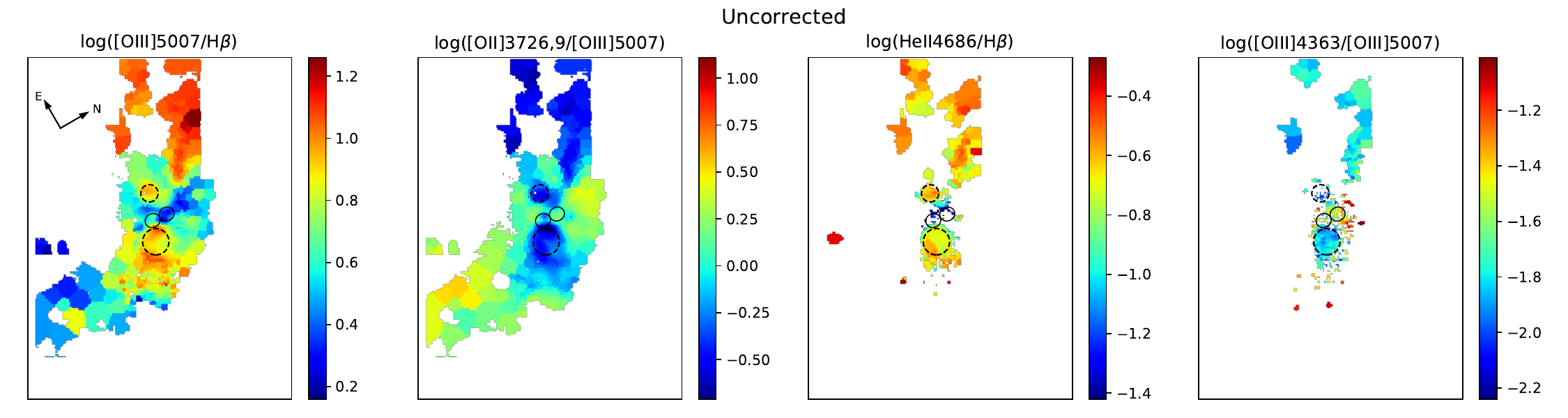}
		\includegraphics[width=\textwidth]{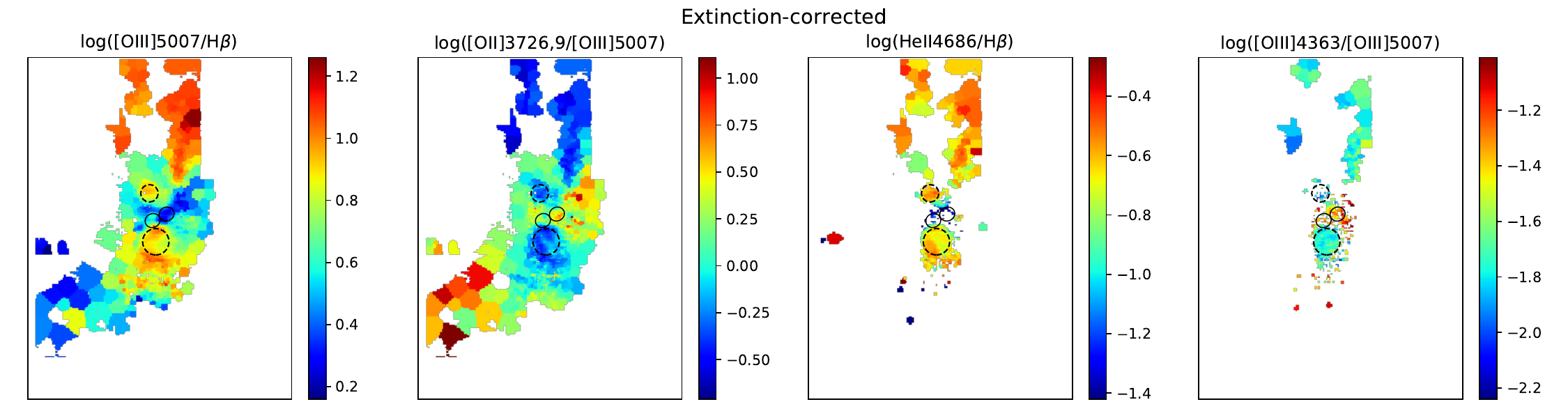}
		\caption{Maps of the [OIII]5007/\hbeta , [OII]3726,9/[OIII]5007, HeII/\hbeta , and [OIII]4363/[OIII]5007 line ratios before ({\em top} row) and after ({\em bottom} row) extinction correction. Each column shows a line ratio. 
		The black dashed circles show the location of the ``outflow regions'', and the black solid circles show the location of the supperbubbles. 
		The line ratios in the central region differ at the locations of the outflows and superbubbles, with the two regions showing opposite line ratio trends. In the ``outflow region'', high line ratios are measured for [OIII]5007/\hbeta ~and HeII/\hbeta . In the superbubbles, high line ratios are measured for [OII]3726,9/[OIII]5007 and [OIII]4363/[OIII]5007. A remarkable feature is that the NE nebula displays similar line ratios as the ``outflow regions''.}
\end{figure*}

\subsection{Emission line ratio maps}

An important goal of this paper is to investigate the origin of the ionized gas that produces the observed emission.
From the numerous emission line maps presented for  Mrk 273, one can obtain line ratio maps, which are crucial in diagnosing the physical conditions of the ionized gas in different regions of the galaxy.
To obtain accurate line ratios, we first account for extinction across the galaxy.
We calculated the reddening $E(B-V)$ in each Voronoi bin from the Balmer decrement of the total \hbeta ~and H$\gamma$ emission line fluxes.
We assumed Case B recombination at $10^4$ K and the \citet{car89} extinction curve and $R_V = 3.1$.
The resulting reddening map is shown in Figure \ref{fig:ebv}.
Relatively strong reddening of $E(B-V) \gtrsim 0.5$ is observed in the the SW nebula as well as the inner nuclear region, while the NE nebula has weaker reddening of $\sim 0.2$.
The median reddening of the galaxy is 0.19.  

We constructed maps of line ratios commonly used as diagnostics before and after extinction correction, and show them in Figure \ref{fig:lr}.
To correct for extinction, line fluxes in each Voronoi bin are calculated using the reddening map.
For Voronoi bins without significant detection of the Balmer lines, the median reddening value across the galaxy is used to calculate the corrected flux.
Extinction corrections lead to a median change in the line ratios of [OIII]5007/\hbeta , [OII]3726,9/[OIII]5007, HeII/\hbeta , and [OIII]4363/[OIII]5007 of -0.01, 0.09, 0.01, and 0.05 dex, respectively.
We note that the change in the numerical values of the line ratios are modest, and the spatial trends seen in the line ratio maps are unchanged by extinction correction.
Numerical values of line ratios in the rest of this subsection refer to the extinction-corrected values.

The NE nebula has high $\log(\mathrm{[OIII]5007}/\hbeta)$ line ratios of $\sim 1$ to 1.2, while these line line ratios are substantially lower in the SW nebula, with values of  $\sim 0.2$ to 0.6.
The $\log(\mathrm{[OIII]5007}/\hbeta)$ line ratios in the nuclear region have a complex structure.
An interesting feature is that regions of high $\log(\mathrm{[OIII]5007}/\hbeta)$ line ratios extend in the northeast and southwest directions, coinciding with the ``outflow region'' (dotted circles), while regions of low line ratios extend in the north and south directions, aligning with the axis of the supperbubbles (circles with solid outlines).

The distribution of the $\log(\mathrm{[OII]3726,9/[OIII]5007})$ line ratio  generally displays opposite trends to that of  $\log(\mathrm{[OIII]5007}/\hbeta)$.
While this is not entirely unexpected because of the inverse dependence on [OIII]5007 of the two line ratios, the different dynamic range of the line ratios indicates that this is also driven by differences in [OII]3726,9.
$\log(\mathrm{[OII]3726,9/[OIII]5007})$ is the highest in the SW nebula, at $\sim 0.5$ to 1, and the lowest in the NE nebula, at $\sim -0.7$ to $-0.3$.
In the nuclear region, the $\log(\mathrm{[OII]3726,9/[OIII]5007})$ line ratios are elevated along the direction of the superbubbles, and are lower in the ``outflow regions''.

The $\log(\mathrm{HeII4686}/\hbeta)$ line ratio exhibits similar trends to $\log(\mathrm{[OIII]5007}/\hbeta)$.
The NE nebula and the ``outflow region'' display enhanced line ratios, while the superbubbles have lower line ratios.
For $\log(\mathrm{[OIII]4363/[OIII]5007})$, higher line ratios are observed along the superbubbles, and lower line ratios are observed in the ``outflow region'' and the NE nebula.

\subsection{Ionization mechanism}\label{subsec:ion}

\begin{figure*}[!htbp]\label{fig:o3hb_o2o3}
	\centering
		\includegraphics[width=\textwidth]{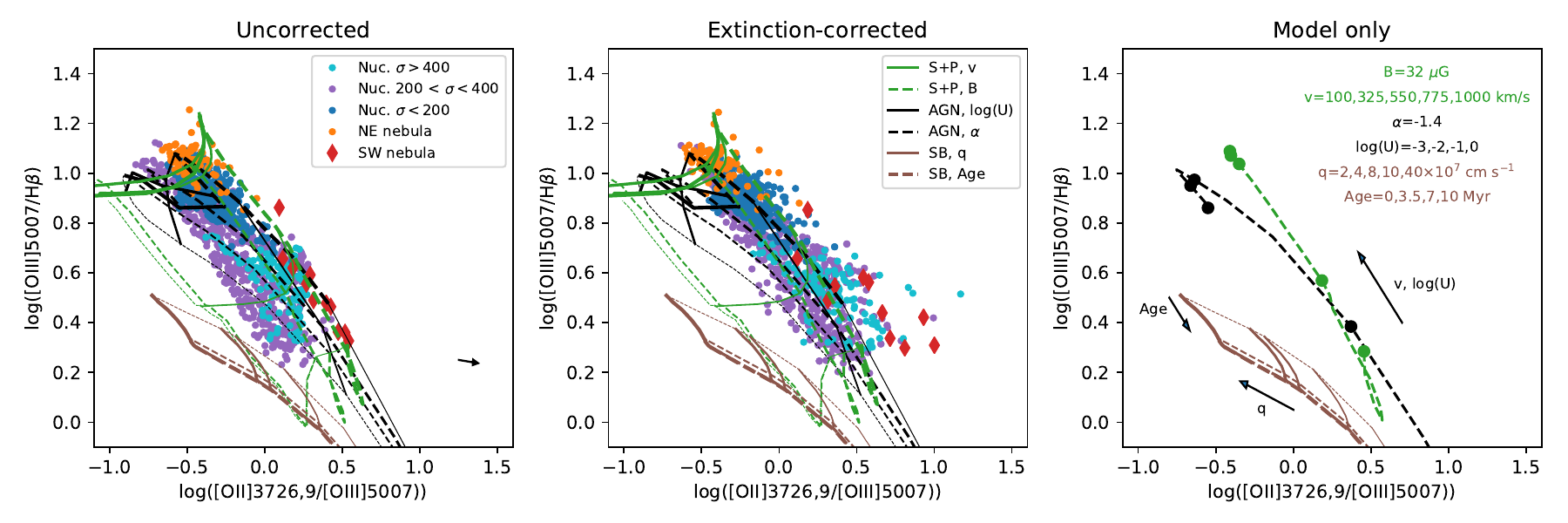}
		\caption{Diagnostic diagram of [OIII]5007/\hbeta ~vs. [OII]3726,9/[OIII]5007. {\em Left}: Line ratios before extinction correction overplotted with model grids. The red diamonds correspond to the SW nebula, the orange points correspond to the NE nebula, the cyan, violet and blue points correspond to the high, medium and low velocity spaxels in the  nuclear region. The green grids show a shock+precursor ionization model with solar metallicity and a pre-shock density of 100 cm$^{-3}$. The solid grid lines correspond to five shock velocity values (100,  325,  550,  775 and 1000 \kmps , thin to thick), while the dashed grid lines correspond to four  magnetic field values (0.001, 1, 20 and 100 $\mu$G, thin to thick). The black grids show a dusty AGN ionization model with solar metallicity and a density of 100 cm$^{-3}$. The the solid grid lines correspond to five values of ionization parameter $\log(U)$ (-4, -3, -2, -1 and 0, thin to thick), while dashed grids lines correspond to four values of the power law index (-2, -1.7, -1.4 and -1.2, thin to thick). The brown grids show a starburst ionization model assuming standard mass loss and continuous star formation, with solar metallicity and a density of 100 cm$^{-3}$. The solid grids lines correspond to five values of ionization parameter $q$ (0.2, 0.4, 0.8, 1 and 4$\times 10^8$, thin to thick), while the dashed grid lines correspond to four values of age (0, 3.5, 7 and 10 Myr, thin to thick). The black arrow in the left panel represents the median magnitude and direction of extinction correction on the line ratios. {\em Middle}: Extinction-corrected line ratios overplotted with model grids. {\em Right}: To allow for easier visualization of the trends of model parameters, we show model sequences without the observed data. The shock+precursor model is shown at a fixed magnetic field strength of $32~\mu$G, and the AGN photoionization model is shown at a fixed power law index of $-1.4$. The shock and AGN models overlap with each other, and both align with the observed line ratios. The starburst model cannot reproduce the observed line ratios in any regions of the galaxy.}
\end{figure*}

\begin{figure*}[!htbp]\label{fig:o3o3_he2hb}
	\centering
		\includegraphics[width=\textwidth]{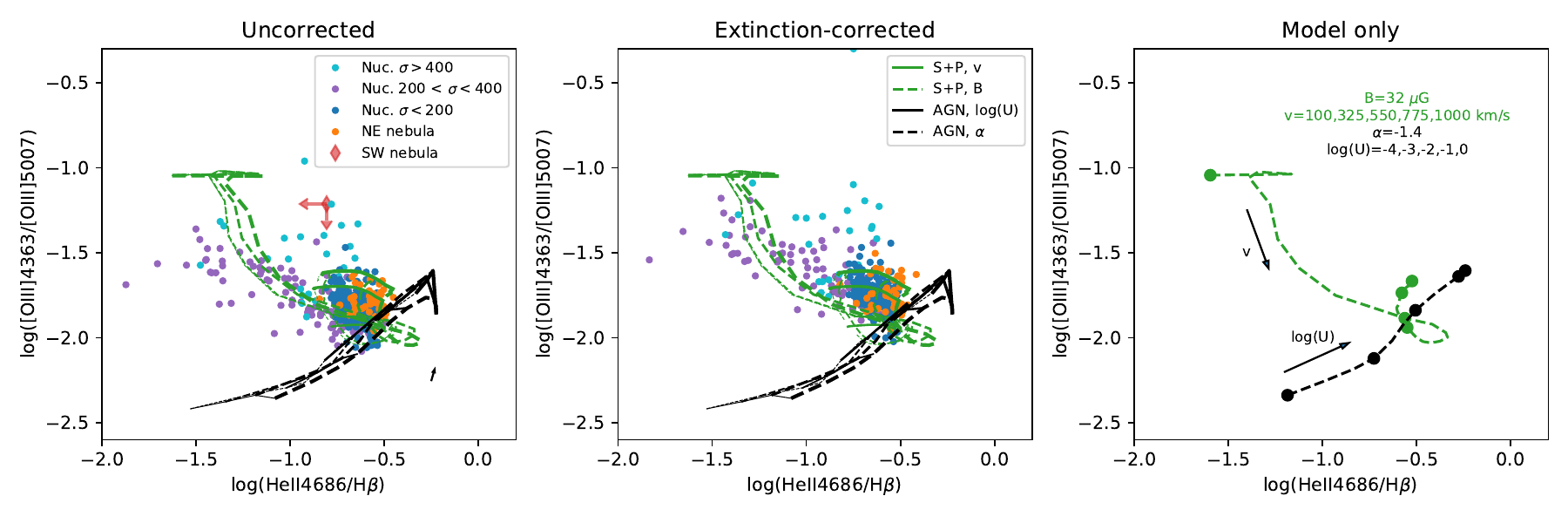}
		\caption{Same as Figure \ref{fig:o3hb_o2o3}, but for [OIII]4363/[OIII]5007 vs. HeII4686/\hbeta line ratios. The shock+precursor model (green grids)  successfully reproduces most of the observed line ratios, while the AGN photoionization model fails to account for most of the spaxels in the galaxy. The red arrow shows the median value of the $3\sigma$ upper limit in the SW nebula.}
\end{figure*}

\begin{figure*}[!htbp]\label{fig:o2o3_o3o3}
	\centering
		\includegraphics[width=\textwidth]{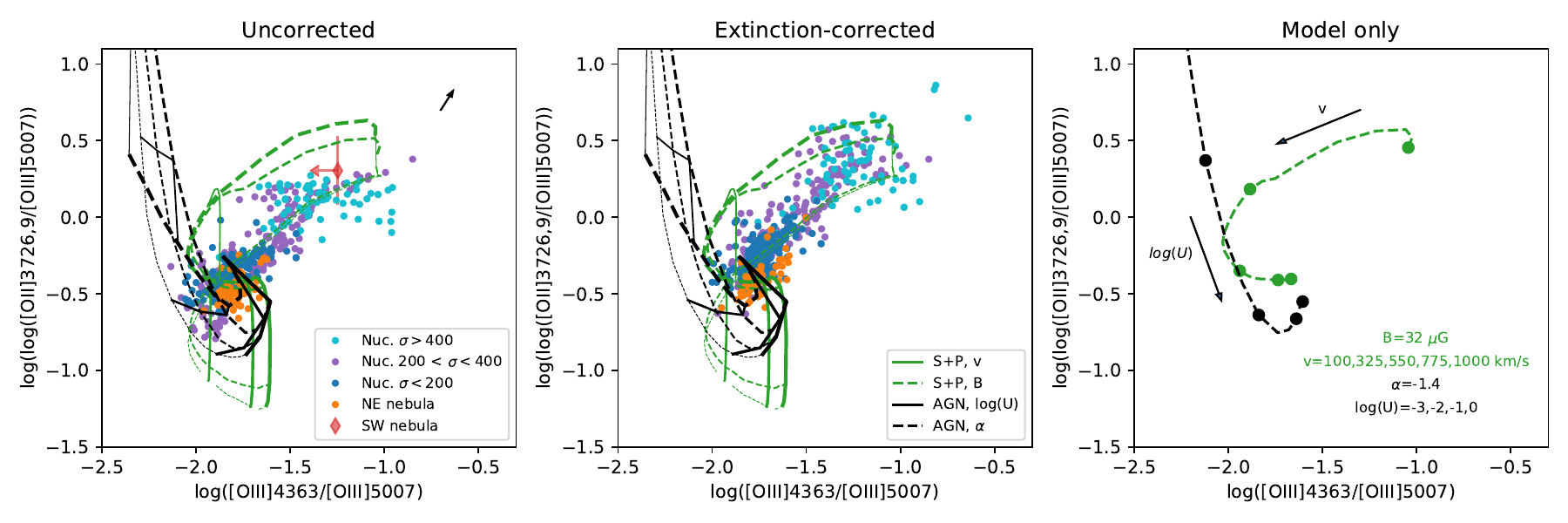}
		\caption{Same as Figure \ref{fig:o3hb_o2o3}, but for [OII]3726,9/[OIII]5007 vs. [OIII]4363/[OIII]5007 line ratios. The shock plus precursor model grids overlap with most of the observed line ratios in all the regions, while the AGN photoionization model only matches a fraction of the spaxels in the NE nebula and the slow spaxels in the nuclear region.}
\end{figure*}

\begin{figure*}[!htbp]\label{fig:o2o3_ne5ne3}
	\centering
		\includegraphics[width=\textwidth]{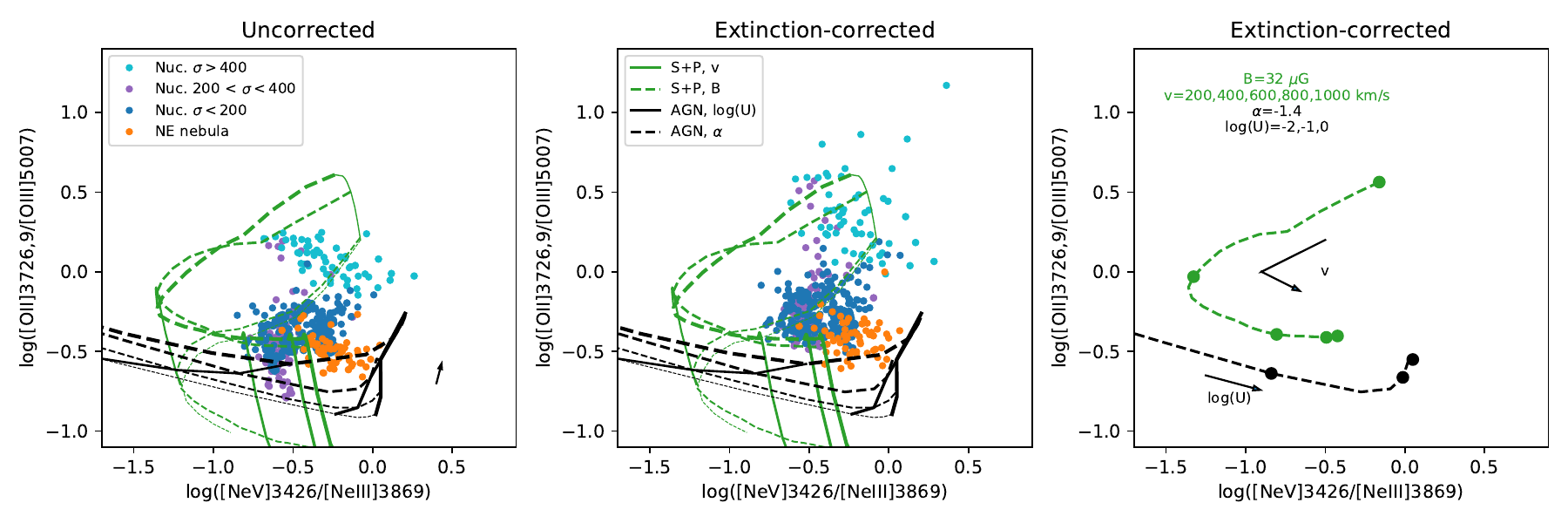}
		\caption{Same as Figure \ref{fig:o3hb_o2o3}, but for [OII]3726,9/[OIII]5007 vs. [NeV]3426/[NeIII]3869 line ratios and different grid values for the shock velocity. The shock and precursor model grids show five shock velocities of 200, 400, 600, 800 and 1000 \kmps , as model grids for shock velocities below $200 ~\kmps$ extend far beyond the axes limits. The shock+precursor model successfully reproduces most of the observed line ratios, while the AGN photoionization model fails to account for most of the spaxels in the galaxy. The red arrow shows the median value of the $3\sigma$ upper limit in the SW nebula.}
\end{figure*}

In this subsection, we use line ratio diagnostic diagrams to investigate the nature of the ionizing sources of the gas in different regions of Mrk 273.
Figures \ref{fig:o3hb_o2o3} to \ref{fig:o2o3_ne5ne3} show the observed line ratios and predictions of ionization models on diagnostic diagrams. 
Line ratios are chosen to be either close in wavelength so that extinction effects are minimized, e.g. HeII4686/\hbeta ~and [OIII]5007/\hbeta , or from the same species so that effects of an unknown metallicity are minimized, e.g. [OII]3726,9/[OIII]5007, [OIII]4363/[OIII5007] and [NeV]3426/[NeIII]3869 \citep{bal81}.
In each diagnostic diagram, we denote line ratios of spaxels in the nuclear region, the NE nebula, and the SW nebula with different colors to analyze the ionization mechanism in each region.
When an emission line is undetected in a region, we show in the diagnostic diagrams the median value of the $3\sigma$ upper limit in the spaxels of that region.
For line ratios in the nuclear region, we further separate the spaxels in the slow, medium, and fast regions according to their velocity dispersion ($\sigma < 200~\kmps$, $200~\kmps < \sigma < 400~\kmps$ and $\sigma > 400~\kmps$), and denote them with different colors.
In the following, we use the IDL Tool for Emission-line Ratio Analysis \citep[ITERA; ][]{gro10} to explore models of different ionization mechanisms and compare them with the observed line ratios.

\subsubsection{Starburst photoionization}

Since ULIRGs are known to have vigorous starburst activity, and Mrk 273 has a SFR of $139 ~\msun ~\mathrm{yr^{-1}}$, we first study the possibility of starburst photoionization driving the observed emission.
We use the continuous star formation model at standard mass loss rate of \citet{lev10}, assuming solar metallicity and an electron density of $100 ~\mathrm{cm}^{-3}$.
The \citet{lev10} model uses the MAPPINGS III code for radiative transfer, and the Starburst99 code of \citet{vaz05}.
The free parameters are the age and ionization parameter $q$ , and they vary between $0-10$ Myr and $2-40 \times 10^7 ~\mathrm{cm~s}^{-1}$, respectively.
The model grids are overplotted in brown with the observed line ratios in Figure \ref{fig:o3hb_o2o3}.
The starburst model grids do not overlap with the observed line ratios in any of the regions of Mrk 273. 
Some of the spaxels in the nuclear regions are the closest to the models, but the predicted model [OIII]5007/\hbeta ~and [OII]3726,9/[OIII]5007 line ratios are at least 0.3 dex lower than observed.
For the line ratios of higher ionization lines,
the starburst model predicts [OIII]4363/[OIII]5007 line ratios of $<-2.4$ and HeII4686/\hbeta ~line ratios of $<-3.3$, while the lowest observed values are $\gtrsim -2.2$ for both line ratios.
The model grids are not shown in these diagnostic diagrams since they reside mostly outside of the range of the plot.
We also explored the starburst model assuming twice solar metallicity or a density of $10 \mathrm{cm^{-3}}$, and it resulted in a larger disagreement between the model predictions and the observed data.
This rules out starburst photoionization in all spatial regions of Mrk 273, consistent with previous findings using the standard BPT diagram \citepalias{rz14, s16}.

\subsubsection{AGN photoionization}

Since Mrk 273 is known to host at least two AGN in its central region, we next explore the possibility of AGN photoionization.
We apply the dusty, radiation pressure dominated AGN model of \citet{gro04}, assuming solar metallicity and an electron density of $100 ~\mathrm{cm}^{-3}$. 
The model assumes a simple power-law radiation field to represent the AGN ionizing spectrum and a constant gas pressure, and includes dust and the effect of radiation pressure on dust.
The free parameters are the AGN power law index ($\alpha$) and ionization parameter ($\log(U)$), which vary between $-2$ to $-1.2$ and $-4$ to 0, respectively.
We overplot the model grids in black with the observed line ratios in Figures \ref{fig:o3hb_o2o3} to \ref{fig:o2o3_ne5ne3}, and in the right panels, plot a model sequence of varying ionization parameter at a fixed power law index of $\alpha=-1.4$ for easier visualization of the model dependence in each line ratio space.
We also explore AGN photoionization models with half or twice solar metallicity, or an electron density of $1000 ~\mathrm{cm}^{-3}$, holding the other parameter constant, and find that these parameters produce a larger disagreement with the observed data.

The AGN model grids reproduce the observed line ratios in all three spatial regions in the [OIII]5007/\hbeta ~vs. [OII]3726,9/[OIII]5007 plane (Figure \ref{fig:o3hb_o2o3}).
The spaxels in the nuclear region occupy the sequence of model line ratios with ionization parameter ranging from $-3$ to 0.
The fast ($\sigma > 400 ~\kmps$) spaxels lie in the grids of $\log(U) \sim -3$, while the slow ($\sigma < 200 ~\kmps$) spaxels reside in those of $\log(U) \sim -2$ to 0.
The medium ($200  ~\kmps < \sigma < 400 ~\kmps$) spaxels occupy the entire range of $\log(U) \sim -3$ to 0.
The spaxels in the NE nebula occupy model grids at high ionization parameters of $\log(U)$ between $-2$ and 0, similar to the slow spaxels in the nuclear region.
The spaxels in the SW nebula reside in model grids at low ionization parameters of $\log(U) \sim -3$.

However, the AGN model grids cannot reproduce the observed line ratios in the nuclear and NE regions in diagnostics plots involving higher ionization lines.
This is the most prominent in 
Figure \ref{fig:o3o3_he2hb}, where the medium and fast spaxels in the nuclear region have high $\log(\mathrm{[OIII]4363/[OIII]5007})$ line ratios of $>-1.5$, which are too high to be fitted by the AGN model even at the highest ionization parameter.
The slow spaxels in the nuclear regions and spaxels in the NE nebula occupy line ratios somewhat closer to the model predictions with $\log(U) \sim -2$, but the majority of the spaxels are still $\sim 0.3$ dex higher than the model in $\log(\mathrm{[OIII]4363/[OIII]5007})$.
Since both the [OIII]4363 and HeII4686 emission lines are undetected in the SW nebula, only upper limits are available in this diagnostic diagram.
We show the median upper limit of the spaxels in the SW region.
The upper limit in HeII4686/\hbeta ~in the SW nebula is consistent with AGN photoionization with $\log(U) \sim -3$.

Similar results are observed in 
Figures \ref{fig:o2o3_o3o3} and \ref{fig:o2o3_ne5ne3}.
The fast and medium spaxels in the nuclear region deviate the most from the AGN photoionization model predictions.
The slow spaxels in the nuclear region and the spaxels in the NE nebula deviate less from the model predictions, but at least half of these spaxels in Figure \ref{fig:o2o3_o3o3} and the majority of these spaxels in Figure \ref{fig:o2o3_ne5ne3} cannot be reproduced by the AGN photoionization model.
We also show the median upper limit of [OIII]4363/[OIII]5007 in Figure \ref{fig:o2o3_o3o3}, and it is consistent with AGN photoinoization with $\log(U) \sim -3$.

Our results rule out pure AGN photoionization in the fast and medium spaxels in the nuclear region.
While the slow nuclear spaxels and the NE spaxels are closer to the predictions of the AGN photoionization model, our results indicate that AGN photoionization can only produce a part of the emission, and a non-trivial alternative source of ionization is responsible for the rest of the emission in these regions.

\subsubsection{Shock ionization}

An alternative ionization mechanism for the gas in Mrk 273 is radiative shocks.
Shocks excitation has been observed in galaxies impacted by galaxy-galaxy interactions and/or outflows \citep[e.g.][]{ric15}.
Past observations of molecular hydrogen lines have shown evidence of shocks in the nuclear region of Mrk 273 \citep{u13, u19}.
We apply the shock+precursor ionization model of \citet{all08}, assuming solar metallicity and a pre-shock electron density of $100 ~\mathrm{cm}^{-3}$.
The model includes emission from both the post-shock region, where gas is shock-excited, and the pre-shock region, where gas is photoionized by the radiation from the pre-shock region.
The free parameters are the shock velocity ($v_\mathrm{s}$) and magnetic field strength ($B$), which vary between $100-1000~\kmps $ and $0.001-100 \mu \mathrm{G}$, respectively.
For a pre-shock density of $100 ~\mathrm{cm}^{-3}$, a magnetic field strength of $B \sim 32 \mu \mathrm{G}$ corresponds to equipartition between thermal and magnetic pressures ($B/n^{1/2} \sim 3-5$; \citealt{all08}).
We also explore a shock-only model and find that  a high pre-shock density of $1000 ~\mathrm{cm}^{-3}$ is required to produce satisfactory agreement with the observed data.
This would imply a post-shock density of $10000 ~\mathrm{cm}^{-3}$ \citep{dop95}, which is an extreme assumption.
Therefore, the shock and precursor model is preferred and shown here.
We overplot the model grids in green with the observed line ratios in Figures \ref{fig:o3hb_o2o3} to \ref{fig:o2o3_ne5ne3}, and in the right panels, plot a model sequence of varying shock velocity at a fixed magnetic field strength of $B = 32 \mu \mathrm{G}$ for easier visualization of the models. 

All four diagnostic diagrams produces consistent results when comparing the observed line ratios with the shock and precursor model grids.
The observed line ratios of spaxels in both the nuclear region and the NE nebula generally reside within the model grids.
The medium velocity spaxels in the nuclear region occupy a wide range of model grids along shock velocity values of $100-1000 ~\kmps$.
A peculiar result is that in the nuclear region, the fast spaxels occupy the low shock velocity grids ($v_\mathrm{s} \lesssim 300 \kmps $), while the slow spaxels occupy the high shock velocity grids ($v_\mathrm{s} \gtrsim 300 \kmps $).
The spaxels in the NE region occupy a similar space in line ratios with the slow spaxels in the nuclear region, in the high $v_\mathrm{s}$ grids.
For the SW nebula, the line ratios in Figure \ref{fig:o3hb_o2o3} and the upper limits in Figure \ref{fig:o2o3_o3o3} agree with the model grids at $v_\mathrm{s} \sim 100-400 \kmps $.
The upper limits in Figure \ref{fig:o3o3_he2hb} provides a stricter constraint of $v_\mathrm{s} \sim 200 \kmps $.

Another test relevant for shock ionization is the energy budget of the observed emission.
The self-consistent model of radiative shocks with precursor preionization presented in \citet{dop17} shows that radiative shocks with $v_\mathrm{s} \gtrsim 110 \kmps $ and a preshock density of $100 ~\mathrm{cm}^{-3}$ can produce emission with \hbeta ~surface brightness of $\sim 7 \times 10^{-15} ~\mathrm{erg~cm^{-2}~s^{-1}~arcsec^{-2}}$.
In Mrk 273, the peak extinction-corrected \hbeta ~surface brightness in the nuclear region, NE nebula and SW nebula are $\sim 10$, $0.15$ and $0.1 \times 10^{-15} ~\mathrm{erg~cm^{-2}~s^{-1}~arcsec^{-2}}$, respectively.
The observed surface brightness is broadly in agreement with the theoretical predictions, indicating that shocks are an energetically feasible ionization mechanism for the gas in the nuclear region, NE nebula, and SW nebula.

\begin{figure}[!htbp]\label{fig:oiii_vs_r}
	\centering
		\includegraphics[width=0.5\textwidth]{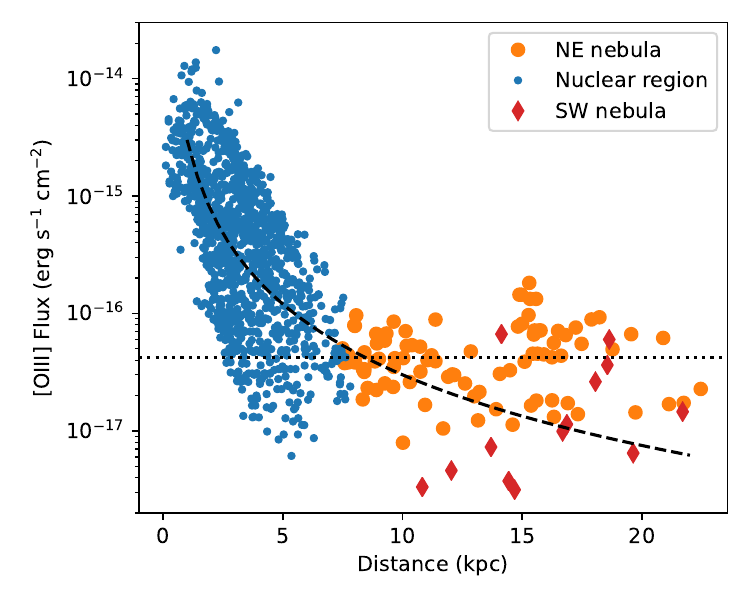}
		\caption{[OIII]5007 flux as a function of distance from the nucleus. The blue points show the spaxels in the nuclear region, orange points the NE nebular and red points the SW nebula. The black dashed line shows a fiducial $r^{-2}$ trend. The horizontal black dotted line shows the median flux in the NE nebula. The [OIII] flux of the NE nebula deviates from the inverse square trend, and displays a relatively flat trend in distance. This is not consistent with a simple AGN photoionization model originating from the nucleus, but is possible to result from in-situ shock excitation.}
\end{figure}

The surface brightness profile of the NE nebula also provides information on its excitation mechanism.
For AGN photoionization, a surface brightness profile declining with distance from the nucleus would be expected, as the ionization photons from the AGN decline as $r^{-2}$.
On the other hand, shock emission is produced in-situ, and can produce strong emission at the location of the shock, independent of its distance from the nucleus.
The observed surface brightness of the NE nebula is not strongly dependent on the distance from the nucleus, with various clumps with similar peak brightness along the NE direction.
In Figure \ref{fig:oiii_vs_r} we plot the extinction-corrected [OIII] flux in each spaxel versus the distance from the nucleus.
While the flux of the spaxels in the nuclear region follows approximately an inverse square trend, the flux of the spaxels in the NE nebula is relatively flat with respect to distance.

Our results indicate that shock and precursor ionization is contributing to the emission in the nuclear region and the NE nebula.
The reversed relation between the measured velocity dispersion of spaxels and the shock velocity implied by the model predictions suggests that a combination of ionization mechanisms may be present.
As starburst models fail to produce the high line ratios observed, the most likely scenario is the co-existence of shock and AGN ionization in the nuclear region and NE nebula.
The larger deviation from the AGN photoionization model predictions of the fast spaxels in the nuclear region suggests a relatively low contribution from AGN photoionization, and a higher contribution from shocks, which is expected from the higher measured velocity dispersion in these spaxels.
The slow spaxels in the nuclear region and the spaxels in the NE nebula have a smaller deviation from the AGN model predictions, suggesting a relatively higher AGN contribution and lower shock contribution than the fast spaxels.
This higher AGN contribution in slow spaxels  skews the line ratios towards the AGN model predictions, which coincide with the fast shock model grids, leading to the reversed relation between the measured velocity dispersion of spaxels and the shock velocity.
Our current observations cannot conclusively determine the contribution of AGN photoionization and shock ionization in the SW region.

\section{Discussion}\label{sec:discussions}

\subsection{Ionized Outflows in Mrk 273} \label{sec:oflw}

\begin{figure*}[!htbp]\label{fig:schem}
	\centering
		\includegraphics[width=\textwidth]{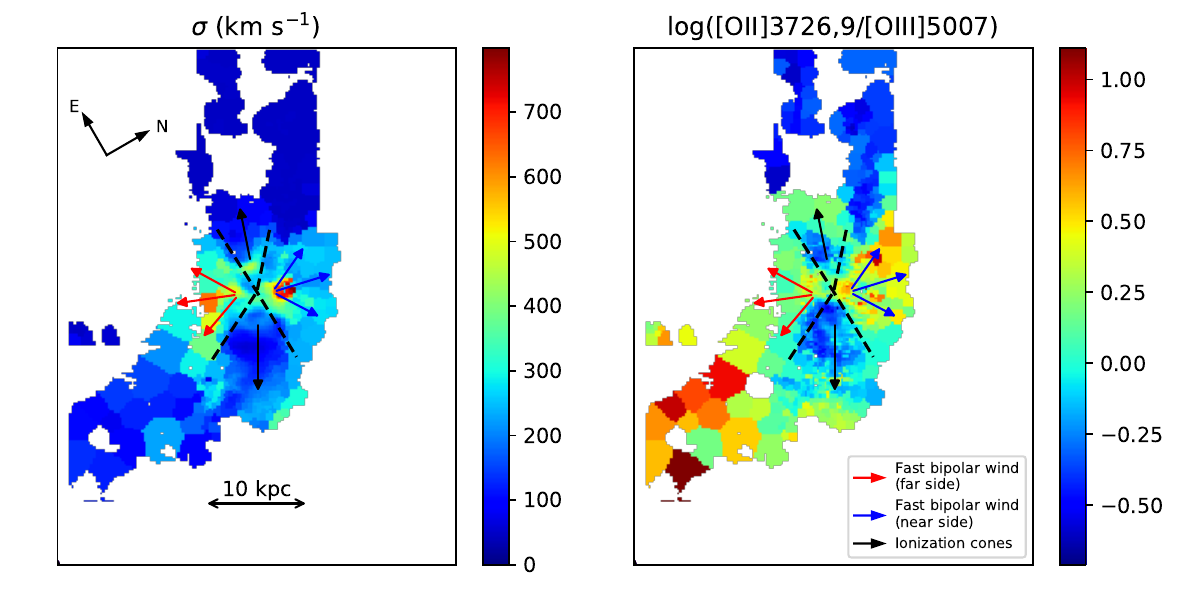}
		\caption{{\em Left}: Velocity dispersion map of ionized gas traced by [OIII]5007 replicated from Figure \ref{fig:kin}. {\em Right}: Extinction-corrected map of the [OII]3726,9/[OIII]5007 line ratio replicated from Figure \ref{fig:lr}. The red and blue arrows show the location and direction of the far and near sides, respectively, of the fast bipolar wind. The bipolar wind is in the N-S direction of the previously reported superbubble in \citetalias{rv13}, but extends $2-3$ times further in radius to over 5 kpc, and has an opening angle of $\sim 90\degr$. The bipolar wind reaches maximum blueshifted or redshifted velocities of up to $1800 ~\kmps$. The black arrows show the location and direction of the ionization cones and/or the slower outflows. The ionization cones encompass the high [OIII] flux ``outflow regions'' defined in \citetalias{rz14}. Line ratios in the ionization cones show a mixing of shock plus precursor excitation and AGN photoionization, with an increased contribution from the latter compared with the rest of the galaxy. The ionized gas in the ionization cones show velocity dispersion of $150-250 ~\kmps$, and bipolar central velocities of $\sim 100 ~\kmps$, suggesting possible outflows relatively perpendicular to the line of sight (see Section \ref{sec:oflw} for discussion). The black dashed lines show the separation between the bipolar wind and the ionization cones, centered at the midpoint between the N and SW AGNs.}
\end{figure*}

An important result of this study is the mapping of the kinematics of the ionized gas outflows in the nuclear region of Mrk 273.
Along the N-S direction of the superbubble reported by \citetalias{rv13}, we detect high velocity gas with velocity dispersion $> 250 ~\kmps$ in two wedge-shaped regions extending in both directions, demarcated by black dashed lines in Figure \ref{fig:schem}.
Moreover, the measurement of $v_{02}$ and $v_{98}$ shows the same wedge-shaped distribution of gas kinematics, with the N side blueshifted and the S side redshifted to maximum velocities up to $1800 ~\kmps$.
The wedges span about 5 kpc in radius and $\sim 90\degr$ in opening angle, and are approximately $2-3$ times more extended than the reported radius of the superbubble and up to a factor of 10 larger in projected area.
This clearly demonstrates a fast, extended, wide-angle, bipolar wind in the N-S direction, which was previously identified as a more localized superbubble.
This illustrates the need of large FOV IFU observations in order to measure the true geometry and extent of outflows.

The fitting of the superbubble by \citetalias{rv13} showed that the outflow center lies between the two AGNs, $0.4''$ from the N and $0.7''$ from the SW, while the uncertainty in the outflow center is $0.3''$.
Around the N AGN, a rotating gas disk has been observed with its rotational axis oriented in the NNW-SSE direction, with a PA of $0\degr$ to $-30\degr$ \citep[e.g][]{dow98,u13,ala18}.
This is consistent with the N-S collimation of the bipolar outflow if it is driven by the N AGN.
On the other hand, a morphological stellar disk has been observed around the SW AGN by imaging, although it is not resolved for kinematic analysis \citep{u13,med14}.
The SW disk is elongated along N-S direction, suggesting a rotational axis in the E-W direction.
If it is indeed a rotating disk around the SW AGN, its orientation is not consistent with the N-S collimation of the bipolar outflow.
Therefore, the most likely origin of the N-S bipolar outflow is the N AGN.

Another intriguing structure in Mrk 273 reported in the literature is the ``outflow regions'' defined in \citetalias{rz14} in the ENE and WSW directions.
These regions were selected from the regions with the brightest [OIII] detected in imaging data, and were measured to have disturbed kinematics in selected apertures in long-slit spectroscopy.
As a result, the properties and origin of these ``outflow regions'', such as whether they are a separate outflow from the N-S wind, were not clearly determined.
The kinematic maps of the ionized gas show two wedge-shaped regions of lower velocity dispersion in approximately the E-W directions and between the N-S winds (demarcated by black dashed lines in Figure \ref{fig:schem}). 
The velocity dispersion of the E-W wedges is the highest closer to the nucleus and decreases with increasing radius.
The ``outflow regions'' are situated at the inner tip of the wedges close to the nucleus, and have elevated velocity dispersion of $150-250 ~\kmps$, compared with the median velocity dispersion of $143 ~\kmps$ across the galaxy. 

While emission line ratio diagnostics reveal contributions from shock plus precursor mixed with AGN photoionization in the entire nuclear region, there is a stronger contribution from AGN photoionization in the slower spaxels, corresponding to the E-W wedges in which the ``outflow regions'' are located.
In fact, the line ratio maps also display the same wedge shaped structures in the E-W and N-S directions.
We replicate the [OII]3726,9/[OIII]5007 line ratio map in Figure \ref{fig:schem} and demarcate the wedges with black dashed lines.
Considering the fact that the ``outflow regions'' are also the brightest regions in [OIII], combined with the stronger AGN photoionization contribution, the E-W wedges likely represent the ionization cones of one of the AGNs in the nucleus.
As discussed above, the rotating disk around the N AGN has an axis in the N-S direction, while that around the the SW is in the E-W direction.
Therefore, the E-W wedges most likely represent the ionization cones of the SW AGN.

The kinematics of the E-W wedges are more ambiguous.
While the E-W wedges have an elevated velocity dispersion of $150-250 ~\kmps$, the gas is not as fast as the N-S bipolar outflow, suggesting it is not the same phenomenon as the N-S outflow.
On the other hand, the central velocity $v_{50}$ displays bipolar kinematics, with the E wedge redshifted to $\sim 100-200 ~\kmps$ and the W wedge blueshifted to $\sim 100 ~\kmps$.
Measurements of $v_{02}$ and $v_{98}$ show maximum redshifted velocity of $500-600~\kmps$ in the E wedge and maximum blueshifted velocity of $300-500~\kmps$ in the W wedge, respectively.
The bipolar kinematics are also observed in the central velocity relative to the stellar velocity ($v_{50} - v_*$, Figure \ref{fig:cont}), where the central velocity of the E-W wedges are shifted from the stellar continuum by $\sim 100$ \kmps.
The motion of the gas is decoupled from that of the stars, so the bipolar kinematics is not purely caused by rotation.

It is possible that the E-W wedges represent a slower gas outflow originating from the SW nucleus travelling along the direction of the ionization cone, as outflows are known to escape from paths of least resistance \citep[e.g.][]{fau12}.
The well-defined collimation of the ionization cones, most clearly demonstrated in the line ratio maps (Figure \ref{fig:schem}), also suggests that they are  fairly perpendicular to the line of sight, which can potentially explain the more modest velocity of the gas in the E-W wedges.
The kinematics could also be produced by the remnants of a past outflow episode from the SW nucleus that has partially dissipated away its velocity.

\subsection{Origin of the NE nebula}

One of the most striking features in the KCWI data is the extended NE nebula detected in multiple emission lines spanning over $20$ kpc in extent. 
The NE nebula displays a hollow morphology, which resembles a evacuated bubble possibly originating from a past outflow event \citep[e.g.][]{rup19}.
A particularly remarkable result from the line ratio diagnostics presented here is that the slow spaxels in the compact, nuclear region and the spaxels in the extended NE nebula reside in a very similar line ratio space in all of the diagnostic diagrams.
Another clear  correlation is that not only do both sets of spaxels have similar kinematics ($\sigma < 200 \kmps $), they are also located in the same projected physical direction from the nuclei.
The slow spaxels in the nuclear region are primarily located in the E-W wedges $\sim 1-5$ kpc ENE and WSW of the nucleus.
The NE nebula is located $\sim 5-15$ kpc ahead of the E wedge, northeast of the nucleus.
The similar line ratios, kinematics, and physical direction suggest a common source of ionization for the E-W wedges and the extended NE nebula.

One possible origin of the NE nebula suggested in the literature is tidal debris photoionized by an AGN \citep{rz14, s16}.
This is potentially supported by the diffuse stellar continuum emission at a similar location NE of the galaxy, which is due to tidal interactions.
Moreover, the quiescent kinematics of the NE nebula is not indicative of an active outflow, which is typically kinematically broadened by turbulence to velocity dispersions of $\gtrsim 200 \kmps$.
As the NE nebula is located along the ionization cone of the SW AGN, in this scenario, it can be tidally disrupted gas photoionized by the SW AGN.
The shock contribution shown by line ratio diagnostics could potentially arise from the collision between gas clouds during tidal interactions.

Another possible origin of the NE nebula is one related to an outflow event. 
This is suggested by the bubble-like morphology of the nebula, which could be due to gas evacuated by an outflow \citep[e.g.][]{rup19}.
The kinematics of the NE nebula is relatively narrow (FWHM $\sim 120 ~\kmps$), with a velocity shift with respect to the stellar continuum of $\sim 100 ~\kmps$.
While the kinematics are not suggestive of an active outflow, the nebula could result from a past outflow event, in which the gas has slowed down and dissipated away part of its turbulence.
Such a scenario is possible given the complex outflow-related kinematics revealed by the KCWI data in the inner 5 kpc.
While the fast bipolar wind is travelling in the N-S direction, and thus unlikely related to the NE nebula, the E side of the E-W wedges is in the same projected direction as the NE nebula.
As the E-W wedges potentially represent the remnants of a past outflow episode originating from the SW nucleus that has dissipated part of its velocity (see Section \ref{sec:oflw} for discussion), and the NE nebula could be an extension of the remnants of the same outflow episode.
Assuming an outflow velocity of $500 ~\kmps$, the outflowing gas can reach 20 kpc in $\sim 40$ Myr.
The shock contribution seen in the NE nebula could be produced when the outflow interacts with the ambient materials and dissipates away its velocity.
Recent theoretical studies have noted that ``cold'' ($\sim 10^4$ K) outflowing gas extending large distances (e.g. $\sim 20$ kpc)
can be produced as a result of a cycle of mixing between the hot and cold gas in a starburst-driven outflow and cooling \citep[e.g.][]{arm16, gri17,gro18,sch20}. 
Such scenarios can potentially provide a mechanism to explain the presence of $10^4$ K gas in the NE nebula at 20 kpc from the galactic center through an outflow episode.
Alternatively, recent simulations have shown that such cold gas can arise if cosmic rays are the primary sources of momentum input to cold clouds \citep{bru20}.

\subsection{Detection of Extended [NeV]3426}

The presence of the high-ionization [NeV]3426 emission line in a galaxy spectrum is commonly used as an indicator of AGN activity in galaxy surveys \citep{gil10, mig13, ver18}.
This is because [NeV] has a high ionization potential of $97$ eV, which has been taken to imply the presence of hard ionizing radiation in the soft X-ray or extreme ultraviolet wavelengths, likely emitted by an AGN.
However, models also predict [NeV] emission from shocks, though only at high velocities \citep{all08}.
In Mrk 273, the [NeV]3426 emission line is detected in both the nuclear region and, remarkably, the extended NE nebula at distances of $> 20$ kpc from the nucleus.
Spatially-resolved detections of [NeV]3426 at such large distances are rare.
The emission in these regions, according to emission line diagnostics, is consistent with mixing of shock excitation and AGN photoionization, and shocks can be responsible for part of the extended [NeV]3426 emission.

\section{Conclusions}\label{sec:conclusions}

We present new large-scale KCWI IFU observations of the ULIRG Mrk 273.
We analyze the morphology, kinematics,  and ionization of the ionized gas in the system.
We examine the properties of the ionized gas in three regions: the nuclear region, the extended NE nebula and the extended SW nebula.
We detect multiple emission lines in the two extended nebulae up to $\sim 20$ kpc from the center of the galaxy. 
Our main findings are summarized below:

\begin{enumerate}
    \item In the center of the galaxy we detect high velocity gas with $\sigma > 250 ~\kmps$ along the direction of the previously-detected bipolar superbubbles, but to a distance of $\sim 5$ kpc, indicating a fast bipolar outflow approximately two to three times more extended the previously-reported superbubbles. 
    
    \item Moderately broad emission with $\sigma \sim 150-250 ~\kmps$ is observed in two wedges in the E-W directions, encompassing the previously reported ``outflow regions'', extending $\sim 5$ kpc in the ENE and WSW directions. 
    
    \item The larger scale extended nebulae have fairly uniform velocity dispersion, $\sigma \sim 100-150 ~\kmps$ in the SW nebula and $\sigma \sim 50 ~\kmps$ in the NE nebula.
    
    \item We detect high ionization [NeV]3426, [OIII]4363, and HeII4684 emission  in the NE nebula, tracing the morphology of the nebula as seen in [OIII]5007 emission. These high ionization lines are not detected in the SW nebula.
    
    \item The ratios of various emission lines within the nuclear region are correlated with the kinematic structures. In the ``outflow region'', high line ratios are measured for [OIII]5007/\hbeta ~and HeII/\hbeta . In the superbubbles, high line ratios are measured for [OII]3726,9/[OIII]5007 and [OIII]4363/[OIII]5007.

    \item Line ratio diagnostics of high ionization emission lines show non-trivial shock plus precursor excitation in the NE nebula and the nuclear region. There appears to be mixing between AGN photoionization and shock excitation in both regions. Stronger shock contribution is observed in the nuclear superbubbles, while relatively strong AGN photoionzation is observed in the ``outflow regions'' and the NE nebula.

    \item The NE ``outflow region'' and NE nebula display similar kinematics and line ratios and are spatially connected, suggesting a common origin.
    
    \item We discuss possible scenarios for the production of the ``cool'' ionized gas in the NE nebula at a distance of $\sim 20$ kpc. These data are useful for simulations of the formation of such extended ionized gas around galaxies.
    
\end{enumerate}

Mrk 273 is a unique source at a critical stage of evolution with potentially powerful feedback in action.
Our results present new observations with 
extensive 
spatial and wavelength coverage of Mrk 273.
We reveal in detail the complex two-dimensional kinematics in the nuclear region as well as in the extended large-scale nebulae.
Maps of multiple high ionization emission lines allow for spatially-resolved diagnostics of ionized gas in the galaxy.
Our results demonstrate the power of IFU observations in uncovering the complex features of ULIRGs and AGN-driven outflows, vital to our understanding of the process of mergers and feedback in the evolution of galaxies.

\vspace{0.2cm}
We thank the referee for their comments, which improved the paper.
We also thank Clive Tadhunter for providing the HST narrow band image of Mrk 273.
ALC acknowledges support from the Ingrid and Joseph W.\ Hibben endowed chair at UC San Diego.
The data presented herein were obtained at the W. M. Keck Observatory, which is operated as a scientific partnership among the California Institute of Technology, the University of California and the National Aeronautics and Space Administration. The Observatory was made possible by the generous financial support of the W. M. Keck Foundation.
The authors wish to recognize and acknowledge the very significant cultural role and reverence that the summit of Maunakea has always had within the indigenous Hawaiian community.  We are most fortunate to have the opportunity to conduct observations from this mountain.

\bibliographystyle{aasjournal}
\bibliography{mybib}

\end{document}